\documentclass[journal]{IEEEtran}
\usepackage{xcolor,soul,framed} 

\colorlet{shadecolor}{yellow}
\usepackage[pdftex]{graphicx}
\graphicspath{{../pdf/}{../jpeg/}}
\DeclareGraphicsExtensions{.pdf,.jpeg,.png}

\usepackage[cmex10]{amsmath}
\usepackage{tabularx}
\usepackage{array}
\usepackage{mdwmath}
\usepackage{mdwtab}
\usepackage{eqparbox}
\usepackage{url}
\usepackage{stackengine} 
\usepackage{epstopdf}
\usepackage{cite} 
\usepackage{caption}
\usepackage{subcaption}
\usepackage{breqn}
\usepackage{mathtools, cuted}
\usepackage{lipsum, color}
\usepackage{hyperref}  
\hypersetup{
	colorlinks=true,
	linkcolor=blue,
	filecolor=blue,      
	urlcolor=blue,
	citecolor=blue,
}

\newcommand{\Vbias}{1.1~}
\newcommand{\PthMin}{-3.4~}
\newcommand{\PthAnalytical}{-4.8~}

\newcommand{\ILMin}{0.94~}
\newcommand{\ILSimulation}{0.9~}
\newcommand{\ILAnalytical}{0.6~}

\newcommand{\ISMaxLessThanPmax}{5.4~}
\newcommand{\ISMaxAtThirthydBm}{12.0~}

\newcommand{\PMaxdBm}{13~}
\newcommand{\PMaxAnalyticaldBm}{14~}

\newcommand{\FinOperatingGHz}{2.06~}
\newcommand{\FinOperatingMHz}{2060~}
\newcommand{\FinOperatingApproxGHz}{2.1~}

\newcommand{\FinOperatingRangeMinGHz}{1.85~}
\newcommand{\FinOperatingRangeMaxGHz}{2.1~}
\newcommand{\FreqOperatingRangeMHz}{250~}
\newcommand{\FreqTuningRange}{0.12~}

\newcommand{\PinSweepMin}{-10~}
\newcommand{\PinSweepMax}{28~}
\newcommand{\FBW}{17}

\newcommand{\PinLowPowerSigdBm}{-30~}

\newcommand{\VaractorModelNumber}{Skyworks SMV1231}
\newcommand{\BiasTeeModelNumber}{Inmet 8800SMF3-06}

\newcommand{\CombinerModelNumber}{Mini-Circuits ZFRSC-42-S+}

\newcommand{\CvValue}{2.0~}
\newcommand{\QvValue}{15}
\newcommand{\ZtxValue}{31~}

\newcommand{\PCBAreacm}{2.2~} 

\begin{document}
%
\title{Reflective Parametric Frequency Selective Limiters with sub-dB Loss and $\mu$Watts Power Thresholds}

\author{ Hussein~M.~E.~Hussein,~\IEEEmembership{Student Member,~IEEE,}
  Mahmoud A. A. Ibrahim,~\IEEEmembership{Student Member,~IEEE,}
  Matteo Rinaldi,~\IEEEmembership{Senior Member,~IEEE,}
  Marvin Onabajo,~\IEEEmembership{Senior Member,~IEEE,}
  and~Cristian~Cassella,~\IEEEmembership{Member,~IEEE} 
}


%


\maketitle

\begin{abstract}
This article describes the design methodology to achieve reflective diode-based parametric frequency selective limiters ($pFSLs$) with low power thresholds ($P_{th}$) and sub-dB insertion-loss values ($IL^{s.s}$) for driving power levels ($P_{in}$) lower than $P_{th}$. In addition, we present the measured performance of a reflective $pFSL$ designed through the discussed methodology and assembled on a FR-4 printed circuit board (PCB). Thanks to its optimally engineered dynamics, the built $pFSL$ can operate around $\sim$\FinOperatingApproxGHz GHz while exhibiting record-low $P_{th}$ (\PthMin dBm) and $IL^{s.s}$ (\ILMin dB) values. Furthermore, while the $pFSL$ can selectively attenuate undesired signals with power ranging from \PthMin dBm to \PMaxdBm dBm, it provides a strong suppression level ($IS>$ \ISMaxAtThirthydBm dB) even when driven by much higher $P_{in}$ values approaching \PinSweepMax dBm. Such measured performance metrics demonstrate how the unique nonlinear dynamics of parametric-based FSLs can be leveraged through components and systems compatible with conventional chip-scale manufacturing processes in order to increase the resilience to electromagnetic interference (EMI), even of wireless radios designed for a low-power consumption and consequently characterized by a narrow dynamic range.
\end{abstract}

\begin{IEEEkeywords}
Parametric Components, Frequency Selective Limiters (FSLs), Auxiliary Generators, Nonlinear Dynamics, Interference Suppression
\end{IEEEkeywords}

%
\IEEEpeerreviewmaketitle

\section{Introduction}

The growing Internet of Things (IoT) is challenging the sharing of the available spectrum by an increasing number of wireless devices that interfere with each other. As a result, the performance of the existing radios keep being affected more and more heavily by strong electromagnetic interference (EMI) lowering the capability to receive the useful information and threatening the integrity of any receivers (RXs), especially when designed for a low-power consumption. For this reason, adaptive RXs with interference filtering capabilities have received growing attention in recent research efforts. In particular, radio frequency (RF) power-sensitive components known as Frequency Selective Limiters (FSLs\cite{Orth1968}) have recently been researched to provide low-power RXs with the unique ability to instinctually suppress received EMI without affecting the capability to simultaneously receive the desired lower-power useful information. Thanks to their power-sensitive electrical response, FSLs can address key operational 
needs, such as a higher resilience to EMI in modern radars or the protection of any critical communication systems from high-power microwave attacks or jamming. Two main classes of FSLs have been previously discussed. One class relies on ferrite-based components\cite{Kotzebue1962, Giarola1979,Jackson1967,Berman1964,Stitzer1983,Adam1993,Stitzer2000,Adam2004,Adam2013,Billeter1968,Gillette2018}, exploiting different types of nonlinear mechanisms in thin-film magnetic materials. While ferrite-based FSLs employing different technologies have been explored, the intrinsic losses associated with any usable thin-film magnetic materials render these FSLs prone to high insertion-loss values for small signals ($IL^{s.s}$ up to 10 dB). In addition, since the magnetic materials used by ferrite-based components cannot be manufactured through fabrication processes compatible to the ones used for Complementary-Metal-Oxide-Semiconductor (CMOS) Integrated Circuits (ICs), ferrite-based FSLs cannot be monolithically integrated with the other active and passive components forming the receive and transmit modules of commercial radios.  

Alternatively, diode-based FSLs \cite{Ramirez2008,Phudpong2009,Ho1961,Wolf1960,Phudpong2007,Hueltes2017,Naglich2016} have been investigated. Such components rely on the electrostatic nonlinearities of solid-state devices and systems to achieve compact FSLs that can be integrated with the other components of commercial and military RF chains to ensure the highest degree of miniaturization. Yet, the fully-passive diode-based FSLs reported to date exhibit much higher power thresholds than those attained by the state-of-the-art (SoA) ferrite-based counterparts \cite{Adam2004}, thus not being adequate to protect the majority of the integrated front-ends used by IoT systems. Just recently, a diode-based FSL relying on an active feedback-loop, using a board-level power detection and an electromechanically tunable cavity resonator has been reported\cite{Yang2020}. Yet, due to the required intense operations allowing to reconfigure its transmission characteristics and regardless of its exceptionally low threshold granted by the use of a sensitive power detector, this reported FSL is not ideal when miniaturized radios with a low-power consumption are needed. 

Driven by the need of developing a new class of passive open-loop phase noise cleaners, known as parametric filters \cite{Cassella2017, Cassella2014}, our group has recently investigated the stability of large-signal periodic regimes in diode-based 2:1 parametric frequency dividers (PFDs \cite{Hussein2020}), even describing a new design methodology to achieve exceptionally low parametric power thresholds ($P_{th}$) in lumped or distributed on-board PFD implementations. By further exploiting the outcomes of this investigation, we recently developed a new battery-less, chip-less and harvester-free sensor tag \cite{Hussein2020SuhbHT}, referred to as subharmonic tag (SubHT), utilizing a 860 MHz diode-based parametric circuit made of lumped off-the-shelf components assembled on a printed substrate. Through the SubHT, we demonstrated that the proper engineering of the dynamics of diode-based parametric circuits permits to achieve extraordinarily low $P_{th}$ values (-18 dBm for an input frequency of 860 MHz) approaching the power threshold attainable by SoA ferrite-based FSLs, even when relying on low quality factor (\textless80) components and on packaged diodes with junction capacitance in the $pF$-range. Furthermore, the demonstration of the SubHT allowed us to unveil a unique dynamical characteristic for parametric circuits. Such circuits, in fact, can exhibit a much lower conversion loss (CL) from an ultra-low power ($\ll$1mW) input signal to a sub-harmonic output signal than the minimum CL obtained by any frequency doubling counterparts relying on the same nonlinear reactance and on the same circuit topology. This powerful feature has been leveraged in this work to achieve a tunable diode-based parametric FSL ($pFSL$) that can exhibit record-low $P_{th}$ (\PthMin dBm for an input frequency close to \FinOperatingApproxGHz GHz) and $IL^{s.s}$ (as low as \ILMin dB) values, along with a significant interference suppression ($IS$) reaching \ISMaxLessThanPmax dB for driving power levels ($P_{in}$) lower than \PMaxdBm dBm, even exceeding \ISMaxAtThirthydBm dB for $P_{in}>$ \PinSweepMax dBm. In the next sections, we will first present the main design criteria and trade-offs to consider in order to achieve $pFSLs$ with low $P_{th}$ and low $IL^{s.s}$. Later, we will discuss the perturbation-based circuit simulation approach that can be followed to predict and optimize the $IS$ value achieved by $pFSLs$ directly from the steady-state circuit simulated electrical response. Finally, we will showcase the measured performance of a $\sim$\FinOperatingApproxGHz GHz diode-based $pFSL$ that we designed and built on a FR-4 printed-circuit-board (PCB) in this work.

\section{Reflective $pFSLs$ – Design Methodology  }
A $pFSL$ is a nonlinear circuit able to instinctually suppress the power flow between its ports when driven by strong RF signals with power ($P_{in}$) exceeding a certain threshold ($i.e.,$  $P_{th}$). In order to do so, $pFSLs$ rely on the nonlinear dynamics of diode-based parametric circuits to trigger a $period$-$doubling$ mechanism leading to abrupt changes in the $pFSLs$’ electrical responses for $P_{in}$ values exceeding $P_{th}$. For low-power RF front-ends in particular, $pFSLs$ are well-suited to enhance resilience to interference, especially when combined with circuit-level linearity improvement methods such as those discussed in \cite{MO1, MO2, MO3}. Two main types of $pFSLs$ can be designed, namely the absorptive and the reflective $pFSLs$. Absorptive $pFSLs$ rely on directional couplers with output and coupled ports terminated on two parametric networks including one or more diodes. The reliance on such a circuit topology renders absorptive $pFSLs$ capable to absorb strong RF signals with power exceeding their $P_{th}$ value. Yet, absorptive $pFSLs$ are characterized by significant $IS$ values only within a narrow range of input power levels centered around an optimal value ($P_{opt}$) that dynamically minimizes the return-loss ($RL$) at the input of their parametric networks. Consequently,  absorptive $pFSLs$ exhibit just negligible $IS$ levels for $P_{in}$ largely exceeding $P_{opt}$. Hence, they are challenging to use in the presence of continuously varying interference power levels, such as the ones frequently captured by mobile radios in practical electromagnetic scenarios, since they leave cascaded components exposed to severe risks of being damaged by EMI with power higher than $P_{opt}$. 
 \begin{figure}[h]
  \begin{center}
  \includegraphics[width=\linewidth]{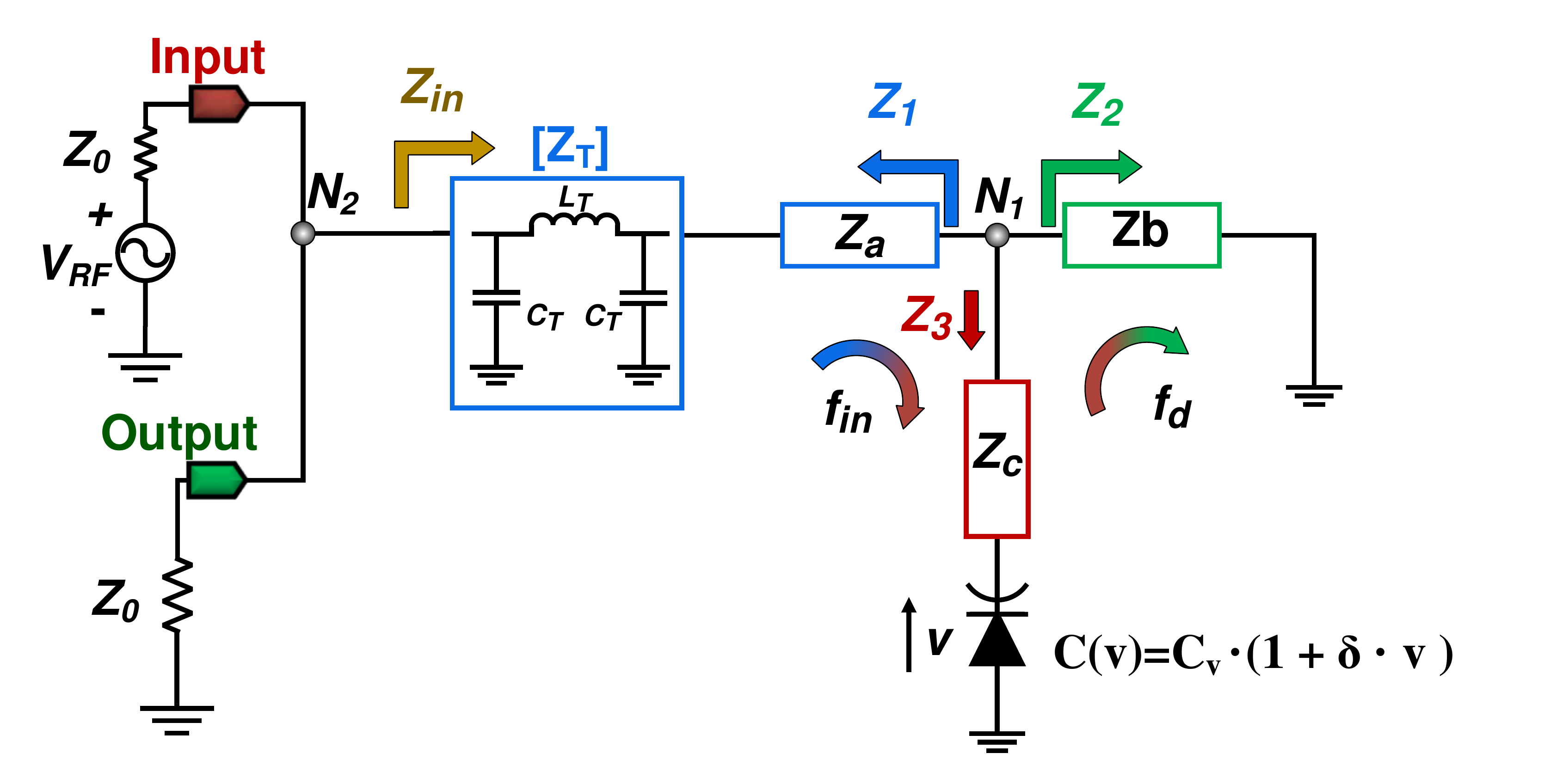}\\
  \caption{ A simplified schematic of a reflective $pFSL$ including three complex impedances ($Z_a$, $Z_b$  and $Z_c$), a quarter-wave transformer ([$Z_T$]), a DC-biased diode and a real termination ($Z_0$) for both its input and output ports. The impedances ($Z_1$, $Z_2$ and $Z_3$) seen from node $N_1$ towards $Z_a$, $Z_b$  and $Z_c$, respectively, are also indicated along with the impedance ($Z_{in}$) seen from the node $N_2$ towards $N_1$. In favor of an easier visualization, no DC-biasing network is shown here.
}\label{generic_schematic}
  \end{center}
\end{figure}
On the contrary, reflective $pFSLs$ exploit the dynamics of diode-based passive circuits to trigger sudden and large increases of $RL$ (at both the $pFSLs$’ input and output ports) in presence of $P_{in}$ values exceeding their $P_{th}$. Reflective $pFSLs$ also exhibit the highest frequency selectivity around an optimal $P_{in}$ value, which depends on the maximum voltage that can be applied across the diode before triggering any periodic transitions between the diode's reverse and forward conduction. Nevertheless, due to their design characteristics, reflective $pFSLs$ can exhibit a high $IS$ even in presence of much stronger interference signals, thus lowering the chances that any receivers can be damaged by EMI of extraordinary high power. However, up to date, a consolidated and systematic design flow for reflective $pFSLs$ is still missing. Such a lack of information, along with the complexity of adapting the $ad$-$hoc$ simulation techniques developed for parametric circuits to the algorithms used by commercial circuit simulators \cite{Cassella2015,Pantoli2008,Ponton2015,Suarez2018,Hernandez2019}, has inhibited the design of reflective $pFSLs$ simultaneously achieving low $P_{th}$ and $IL^{s.s}$ values.

In the most simplistic representation, a reflective $pFSL$ can be seen as a two-port passive network including one diode characterized by a biased capacitance $C_v$ and a corresponding tuning range $\delta$ (see Fig.~\ref{generic_schematic}), together with a set of components forming a stabilization network for the large signal periodic regimes excited in the circuit by $P_{in}$. Without any loss of generality, we can assume any reflective $pFSLs$ to be representable through a T-network topology including the adopted diode, three one-port complex impedances ($Z_a$ , $Z_b$  and $Z_c$) and a quarter-wave transformer (labeled as [$Z_T$]) with bandwidth centered around the value ($f_{in}^{opt}$) of the input frequency ($f_{in}$, corresponding to a natural frequency labeled as $\omega_{in}$) at which the minimum $P_{th}$ is desired. While the transformation stage can be implemented through any existing lumped or distributed circuit topologies, a third-order lumped implementation is assumed here in favor of a simpler analytical treatment. Also, such a two-port network, uniquely identified by an inductance ($L_T$) and a capacitance ($C_T$) setting the equivalent characteristic impedance $(Z_{tx}=\sqrt{ (L_T/C_T )})$, plays a key role to simultaneously achieve the lowest possible $P_{th}$ and $IL^{s.s}$ values in $pFSLs$, as it will be clear in the following section. Furthermore, to match the most frequent operational scenario, the same termination ($Z_0$, equal to 50 $\Omega$) can be considered for the input and output ports. As the $P_{th}$ exhibited by $pFSLs$ needs to be as low as possible to ensure that even low power RXs characterized by a small dynamic range can be protected from EMI, it is crucial to find the optimal design specifications for [$Z_T$], $Z_a$, $Z_b$  and $Z_c$  (from now on labeled together as $Z_{T,a,b,c}$) allowing to minimize the achievable $P_{th}$. By relying on the same analytical methodology used to find the $P_{th}$ of PFDs in our recent theoretical investigation \cite{Hussein2020}, the $P_{th}$ of the reflective $pFSL$ shown in Fig.~\ref{generic_schematic} can be found as: 

\begin{figure*}[t]
\begin{subfigure}{0.5\linewidth}
    \centering
    \caption{}
    \includegraphics[width=1.0\textwidth]{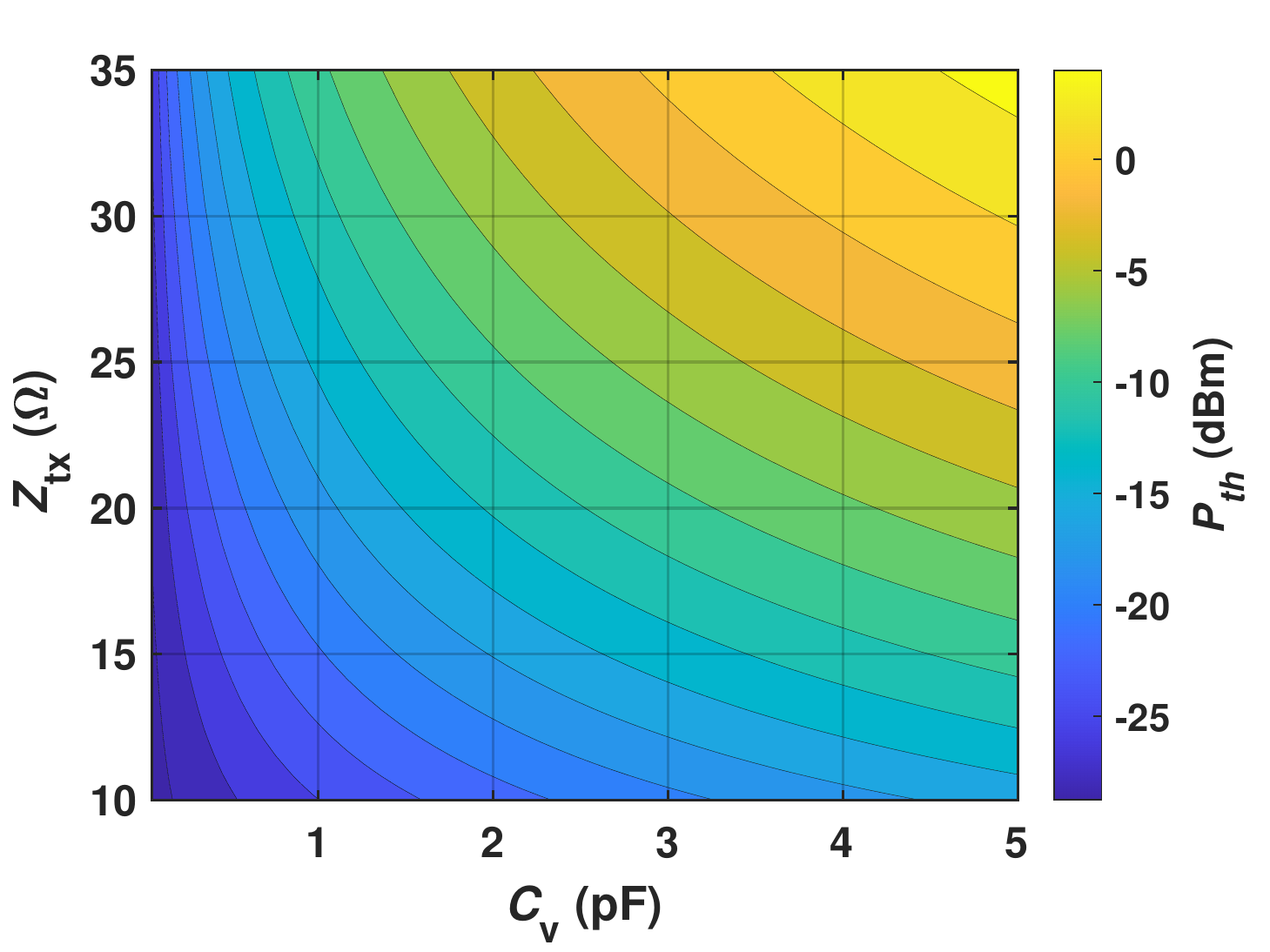} \label{Pth_Vs_Cv_Ztx_contour}
\end{subfigure}
\begin{subfigure}{0.5\linewidth}
    \centering
    \caption{}
    \def\big{\includegraphics[width=1.0\textwidth]{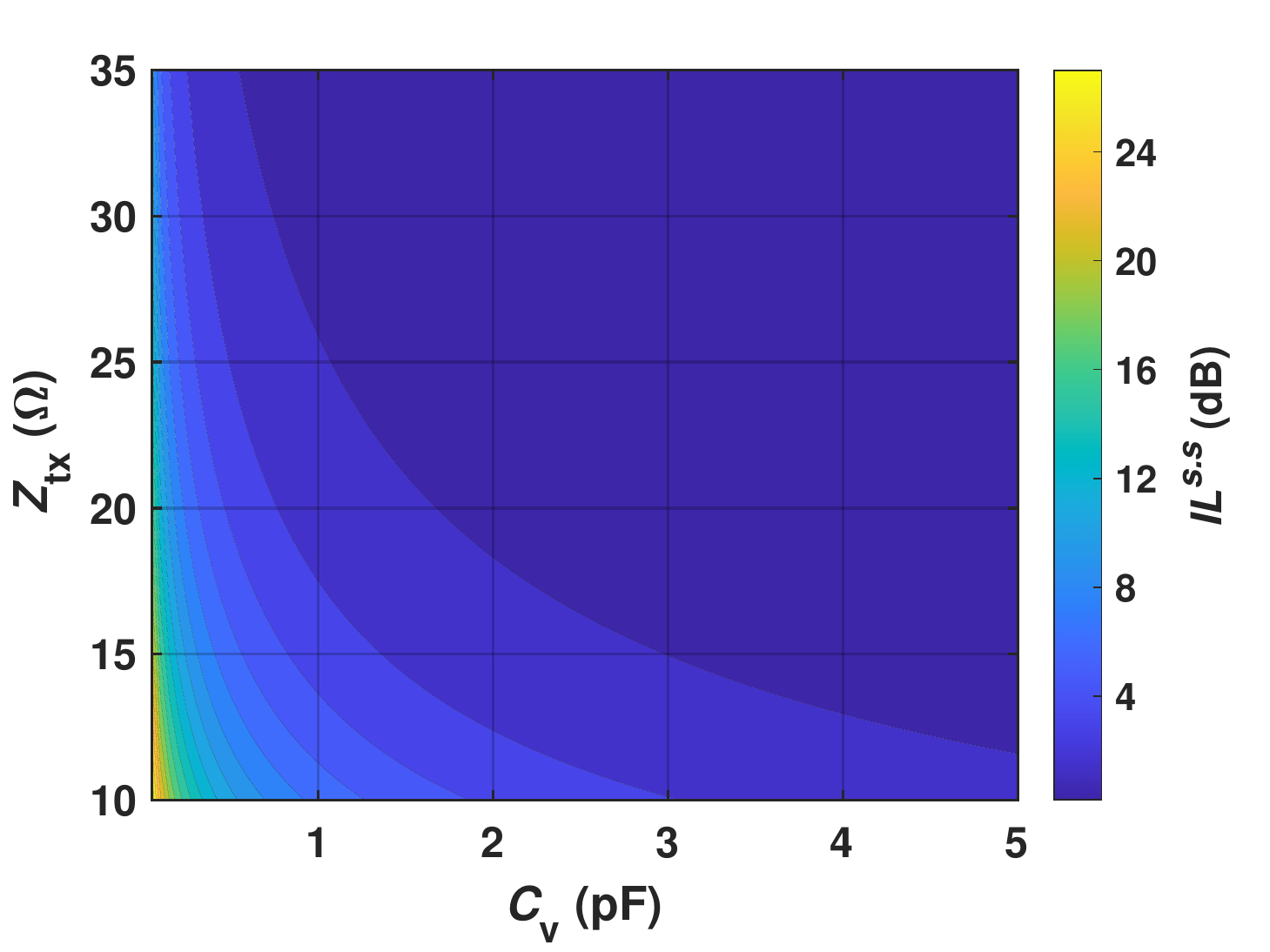}}  
    \def\little{\includegraphics[width=1.6in]{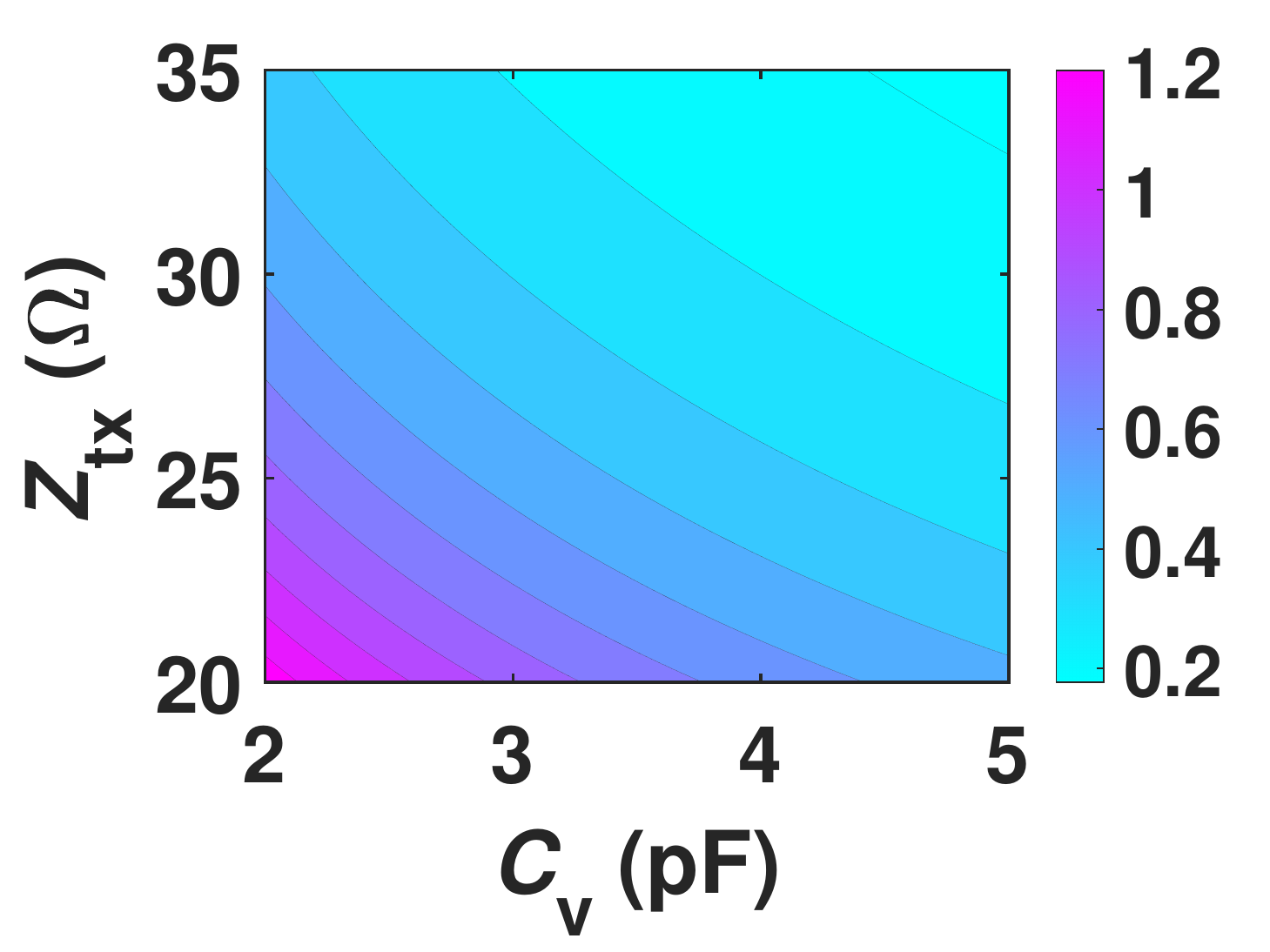}}
    \def\stackalignment{r}  
    \topinset{\little}{\big}{12pt}{48pt}   
    \label{IL_Vs_Cv_Ztx_contour}
\end{subfigure}
\caption{ Minimum $P_{th}$ value (in dBm) computed through \eqref{eq_Pth_Approx} in (a) and minimum $IL^{s.s}$ value (in dB) computed through \eqref{eq_IL_final} in (b) attainable by the reflective $pFSL$ shown in Fig.~\ref{generic_schematic} vs. $C_v$ and $Z_{tx}$. For both a) and b) we assumed $Z_0$ = 50 $\Omega$, $f_{in}^{opt}$=\FinOperatingApproxGHz GHz, $Q_v$=\QvValue, $\delta$=0.4 and the validity of the resonant conditions minimizing $P_{th}$. }
\label{Cv_Ztx_contour}
\end{figure*}


\begin{equation}\label{eq_Pth}
    P_{th}=\frac{|V_{th}|^2}{8 Z_0 } = \frac{C^4_{v}}{2Z_0 \delta^2 } \left|\frac{G^{({\omega}_d)}_{eq} G^{({\omega}_{in})}_{eq} {\omega}^2_{in} }{ \left( Z^{({\omega}_d)}_{1} +  Z^{({\omega}_d)}_{2} \right)  Z^{({\omega}_{in})}_{2} }\right|^2 
\end{equation}

\noindent
In \eqref{eq_Pth}, $V_{th}$ is the voltage level of the input generator with characteristic impedance equal to $Z_0$ and power available equal to $P_{th}$. Also, $Z_1^{\omega_d}$ and $Z_2^{\omega_d}$ represent the equivalent impedances seen at the half of the driving frequency ($f_d$, equal to $f_{in}/2$ and corresponding to a natural frequency labeled as $\omega_d$) from node $N_1$ (see Fig.~\ref{generic_schematic}) towards $Z_a$  and $Z_b$  respectively, whereas $Z_2^{\omega_{in}}$ is the equivalent impedance at $f_{in}$ seen from $N_1$ towards $Z_b$. Further, $G_{eq}^{\omega_{in}}$ and $G_{eq}^{\omega_{d}}$ [see \eqref{eq_G_omega_in} and \eqref{eq_G_omega_d} ] are complex functions of $Z_1^{\omega_{in}}$, $Z_2^{\omega_{in}}$, $Z_3^{\omega_{in}}$ and of the impedances $Z_1^{\omega_d}$, $Z_2^{\omega_d}$ and $Z_3^{\omega_d}$, where $Z_1^{\omega_{in}}$, $Z_2^{\omega_{in}}$ and $Z_3^{\omega_d}$ are the impedances seen at $f_{in}$ and $f_d$ from $N_1$ towards $Z_a$, $Z_b$  and $Z_c$. 
\begin{equation}\label{eq_G_omega_in}
    G^{({\omega}_{in})}_{eq}=Z^{({\omega}_{in})}_{2}  Z^{({\omega}_{in})}_{3}  +Z^{({\omega}_{in})}_{1} ( Z^{({\omega}_{in})}_{2} + Z^{({\omega}_{in})}_{3}) 
\end{equation}
%
\begin{equation}\label{eq_G_omega_d}
    G^{({\omega}_{d})}_{eq}=Z^{({\omega}_{d})}_{2}  Z^{({\omega}_{d})}_{3}  +Z^{({\omega}_{d})}_{1} ( Z^{({\omega}_{d})}_{2} + Z^{({\omega}_{d})}_{3}) 
\end{equation}
By replacing \eqref{eq_G_omega_in} and \eqref{eq_G_omega_d} in \eqref{eq_Pth}, it is straightforward to notice that the resulting $P_{th}$ expression is a function of all the impedances in the circuit ($i.e.$, $Z_{T,a,b,c}$, $Z_0$ and the diode impedance) and of the first two coefficients of the linearized capacitance $vs.$ voltage characteristic of the biased diode (see Fig.~\ref{generic_schematic}). Moreover, since all the impedances in  \eqref{eq_Pth}, \eqref{eq_G_omega_in} and \eqref{eq_G_omega_d} can be extracted through linear algorithms in any circuit simulators, their synthesis can be numerically tackled aiming at the minimization of $P_{th}$ without running time consuming nonlinear perturbation-based simulations, such as the ones available to investigate the steady-state large-signal operation of parametric circuits. Nevertheless, similarly to what we showed for PFDs in \cite{Hussein2020}, the inspection of \eqref{eq_Pth} after simplifying $G_{eq}^{\omega_{in}}$ and $G_{eq}^{\omega_{d}}$ with their corresponding expressions in \eqref{eq_G_omega_in} and \eqref{eq_G_omega_d} permits to realize that the minimum $P_{th}$ at $f_{in}^{opt}$ can be attained by ensuring that four resonant conditions are satisfied: i) $Z_1^{(\omega_{in})}$+$Z_3^{(\omega_{in})}$ must be in series resonance at $f_{in}^{opt}$ with the lowest real part possible ($R_p$); ii) $Z_2^{(\omega_d)}$+ $Z_3^{(\omega_d)}$ must be in series resonance at $f_{in}^{opt}$/2 with the lowest real part possible ($R_d$); iii) $Z_2^{(\omega_{in})}$ must be in parallel resonance at $f_{in}^{opt}$ with the highest real part possible; iv) $Z_1^{(\omega_d)}$ must be in parallel resonance at $f_{in}^{opt}$/2 with the highest real part possible. In order to satisfy these resonance conditions through the use of a minimum number of lumped components, both $Z_a$  and $Z_b$  can be synthesized with parallel $LC$ resonators, resonating at $f_{in}^{opt}$/2 and $f_{in}^{opt}$ and relying on inductors (capacitors) with inductance (capacitance) $L_a$ ($C_a$) and $L_b$ ($C_b$) respectively. Furthermore, $Z_c$ can be synthesized with one inductor whose inductance ($L_c$) is strategically selected to ensure the simultaneous validity of the first two resonant conditions, given the selected $L_a$ and $L_b$ values. Thus, when the resonant conditions mentioned above are satisfied, \eqref{eq_Pth} can be simplified as:

\begin{equation}\label{eq_Pth_simplified}
    P_{th}  = \frac{C^4_{v}}{2Z_0 \delta^2 } 
    \left ( R_p R_d {\omega}^2_{in,opt} \right )^2
\end{equation}

\noindent where $\omega_{in,opt}$ is equal to 2$\pi$$f_{in}^{opt}$ and $R_p$ can be expressed in terms of $Z_{tx}$ [see \eqref{eq_Rp}] when assuming for simplicity [$Z_T$] to be lossless. 

\begin{equation}\label{eq_Rp}
    R_{p}  = R_s + \frac{2 Z^2_{tx}}{ Z_0 }
\end{equation}

\noindent
In \eqref{eq_Rp}, $R_s$ captures the ohmic losses of both $Z_a$ and $Z_c$ along with the ones of the diode. Also, in the derivation of \eqref{eq_Pth_simplified} we neglected the impact of $Z_2^{\omega_{in}}$ and $Z_1^{\omega_d}$ in favor of a simplified analytical treatment and an easier visualization of the main features determining $P_{th}$ in reflective $pFSLs$. The validity of this approximation is granted by the fact that these two impedances, which are synthesized by two notches, just allow to isolate the signals at $f_{in}^{opt}$/2 and $f_{in}^{opt}$ in dedicated meshes of the circuit. From \eqref{eq_Pth_simplified} and \eqref{eq_Rp} it is evident how the transformation stage plays a key role in lowering $R_p$ with respect to the value it would have ($i.e.$, $R_s$+$Z_0/2$) if the resonant conditions were satisfied without using such a transformation stage. Therefore, the adoption of [$Z_T$] introduces fundamental means to reduce the lowest $P_{th}$ that can be attained by reflective $pFSLs$. It is useful to simplify \eqref{eq_Pth_simplified} by expressing $R_s$ and $R_d$ in terms of the quality factor of the diode ($Q_v$, relative to an $f_{in}$ value equal to $f_{in}^{opt}$) and the ones ($Q_1$ and $Q_2$) respectively exhibited at $f_{in}^{opt}$ and $f_{in}^{opt}$/2 by $Z_1^{\omega_{in}}$+$Z_3^{\omega_{in}}$ and $Z_2^{\omega_d}$+$Z_3^{\omega_d}$. Also, for $pFSLs$ operating in the Ultra-High-Frequency (UHF) range, like the prototype demonstrated in this work, we can assume $Q_1$ to be equal to $Q_2$ ($i.e.$, $Q_{1,2}$ = $Q_l$, being $Q_l$ dependent on the technology of the adopted passive components) and to be significantly higher than $Q_v$. This simplification is particularly valid when relying on diodes with wide tuning ranges, such as any available hyperabrupt varactors\cite{Tiggelman2009}. Consequently, $R_s$ can be considered equal to $R_d$ and can be simplified as follows:

\begin{equation}\label{eq_Rs}
    R_{s} =  R_d =  \frac{1}{ \omega_{in,opt} C_v } \left ( 
    \frac{1}{Q_l}+\frac{1}{Q_v}
    \right ) \approx
    \frac{1}{ \omega_{in,opt} C_v Q_v }
\end{equation}

By using \eqref{eq_Rs} in \eqref{eq_Pth_simplified}, $P_{th}$ can then be rewritten as:

\begin{equation}\label{eq_Pth_Approx}
    P_{th}=\frac{(Z_0 + 2 C_v Q_v Z_{tx}^2 \omega_{in,opt} )^2}{2 Q_v^4 Z_0^3 \delta^2  } 
\end{equation}

\noindent
Equation \eqref{eq_Pth_Approx} gives us the opportunity to estimate the minimum $P_{th}$ value that can be attained by reflective $pFSLs$ for any given $f_{in}^{opt}$ of interest and for a chosen diode’s characteristics. As an example, we report a contour-plot capturing $P_{th}$ $vs.$ $Z_{tx}$ and $C_v$ [see \eqref{eq_Pth_Approx}] in Fig.~\ref{Pth_Vs_Cv_Ztx_contour}, which was analytically derived when assuming: i) the use of an hyperabrupt varactor (i.e., $\delta$ $\sim$ 0.4 and $Q_v$ $\sim$ \QvValue) with $C_v$ ranging from 50 fF to 5 pF, and ii) an $f_{in}^{opt}$ of \FinOperatingApproxGHz GHz (in line with our experimental demonstration).


As evident from Fig.~\ref{Pth_Vs_Cv_Ztx_contour}, reflective $pFSLs$ can achieve low $P_{th}$ values ($\ll$0 dBm) through the strategic adoption of diodes with wide tuning range and low capacitance, along with $\lambda$/4 transformation stages characterized by the minimum realizable characteristic impedance. 
Nevertheless, since $pFSLs$ also need to exhibit the lowest possible insertion loss for power levels that are lower than $P_{th}$, the selection of $Z_{T,a,b,c}$ must also be made to ensure a minimum $IL^{s.s}$. In order to estimate the $IL^{s.s}$ value of the $pFSL$ shown in Fig.~\ref{generic_schematic}, we can extract the small-signal scattering parameter ($S_{21}$) for the transmission at $f_{in}^{opt}$ from the $pFSL$'s input port to the $pFSL$'s output port, after linearizing the capacitance $vs.$ voltage characteristic of the diode around $V_{DC}$. The expression of $IL^{s.s}$ in dB is provided in \eqref{eq_IL}.

\begin{equation}\label{eq_IL}
    IL^{s.s.}  = -20 log_{10}(S_{21})= -20 log_{10} 
    \left |
    \frac{
    Z^{({\omega}_{in,opt})}_{in} 
    }{
    Z^{({\omega}_{in,opt})}_{in} +
    \frac{Z_0}{2}
    }
    \right |      
\end{equation}

\noindent
In \eqref{eq_IL}, $Z_{in}^{\omega_{in,opt}}$ is the impedance seen at $f_{in}^{opt}$ from the circuit node $N_2$ (see Fig.~\ref{generic_schematic}) towards $N_1$ and its value is almost independent of $P_{in}$ for $P_{in}$\textless$P_{th}$. Upon validity of the same resonance conditions that minimize $P_{th}$ and $Z_{in}^{(\omega_{in,opt})}$, it can be found that:

\begin{equation}\label{eq_Zin}
    Z^{({\omega}_{in,opt})}_{in}  =
    \frac{ Z_{tx}^2 }{ R_s }
\end{equation}

\noindent
Equation \eqref{eq_Zin} allows to simplify $IL^{s.s}$ [\eqref{eq_IL}] as follows: 

\begin{equation}\label{eq_IL_simplified}
    IL^{s.s.}  = -20 log_{10} 
    \left |
    \frac{ 2 Z_{tx}^2 
    }{
    R_s Z_0 +2 Z_{tx}^2
    }
    \right |      
\end{equation}

\noindent
By replacing \eqref{eq_Rs} in \eqref{eq_IL_simplified}, $IL^{s.s}$ can be finally rewritten as shown in \eqref{eq_IL_final}.
\begin{equation}\label{eq_IL_final}
    IL^{s.s.}  = -20 log_{10} 
    \left |1-
    \frac{  Z_0 
    }{
      Z_0 +2 C_v Q_v Z_{tx}^2 \omega_{in,opt}
    }
    \right |      
\end{equation}

Equation \eqref{eq_IL_final} enables us to assess the minimum $IL^{s.s}$ value that can be attained by reflective $pFSLs$, given a selected $f_{in}^{opt}$ value and based on the DC-biased capacitance of the selected diode. As an example, a contour-plot capturing $IL^{s.s}$ $vs.$ $Z_{tx}$ and $C_v$ is displayed in Fig.~\ref{IL_Vs_Cv_Ztx_contour}, which was derived analytically while assuming the same $\delta$, $C_v$, $f_{in}^{opt}$ and $Z_0$ values used during the derivation of Fig.~\ref{Pth_Vs_Cv_Ztx_contour}. By comparing \eqref{eq_IL_final} with \eqref{eq_Pth_Approx} and Fig.~\ref{Pth_Vs_Cv_Ztx_contour} with Fig.~\ref{IL_Vs_Cv_Ztx_contour}, an important design trade-off between the desired $P_{th}$ and $IL^{s.s}$ values can be identified. While relying on high-$Z_{tx}$ values allows to reduce $IL^{s.s}$, it also determines an increase of $P_{th}$ that can be unacceptable unless $C_v$ values in the $fF$-range were used. Since any board-level $pFSLs$, such as the one we designed and built in this work, can only leverage $C_v$ values close to 1 pF or higher, there exists a fundamental limit for the minimum $P_{th}$ that can be obtained while preserving a low $IL^{s.s}$. Furthermore, $Z_{tx}$ cannot be made arbitrarily large without rendering [$Z_T$] severely affected by electrical loading, thus also degrading $IL^{s.s}$. This inevitably restraints the pool of usable diodes as it limits the maximum acceptable $R_s$ and consequently the minimum $Q_v$ based on the maximum $IL^{s.s}$ that can be tolerated. Therefore, the strategic selection of the diode and of the other components forming $Z_a$ and $Z_c$  is critical to make sure that the lowest $IL^{s.s}$ can be attained without requiring a large $Z_{tx}$ that would compromise the achievable $P_{th}$-value. Moreover, Fig.~\ref{Cv_Ztx_contour} provides useful means to assess the performance that would be achieved if $pFSLs$ were designed and built directly on-chip through any available CMOS technologies. In such a scenario, thanks to the availability of both capacitors and diodes with capacitance in the $fF$-range, any integrated reflective $pFSLs$ would be able to simultaneously rely on extraordinarily low $C_v$ and high $Z_{tx}$ values ($>$1 k$\Omega$), thus enabling much lower $P_{th}$ values ($<$ -20 dBm) than possible with board-level counterparts given a targeted $IL^{s.s}$ value. 

\subsection{Evaluation of the maximum $P_{in}$ value for a preserved frequency selectivity}
While the achievement of low $P_{th}$ and $IL^{s.s}$ values is certainly crucial, another metric to consider during the design of any $pFSLs$ is the maximum $P_{in}$ value ( $ P_{max} = \sqrt{8Z_0 V_{max}}$, where $V_{max}$ is the corresponding peak voltage at $f_{in}^{opt}$) at which a frequency selective attenuation at $f_{in}^{opt}$ is preserved. In particular, with regards to reflective $pFSLs$, the existence of a finite $P_{max}$ is due to periodic transitions from reverse to forward conduction that any diodes undergo when the voltage across their terminals exceeds the sum of the diode's DC bias and built-in ($V_{bi}$) voltages. Due to these transitions, the diode's resistance ($R_v$) undergoes a sudden increase as $P_{in}$ is made larger than $P_{max}$, leading to a progressive reduction of the parametrically generated negative resistance at $f_{d}$ responsible for the rising of the sub-harmonic oscillation in the circuit and, consequently, to a degradation of the $pFSL$ performance at $f_{in}^{opt}$. As a first-order of approximation, $P_{max}$ can be found through a straightforward analysis of the circuit shown in Fig.~\ref{generic_schematic} based on transmission matrices [see \eqref{eq_P_max}] when assuming that $Z_{a}$, $Z_{b}$ and $Z_{c}$ satisfy the resonant conditions minimizing $P_{th}$ and when neglecting (for simplicity) any quadratic or cubic nonlinearities of the diode. 

\begin{equation}\label{eq_P_max}
P_{max}  \approx  \frac{
(V_{DC}+V_{bi})^2 \left( 2 C_v Q_v \omega_{in,opt} Z_{tx}^2+Z_0 \right)^2}
{8 (Q_v)^2 Z_0 Z_{tx}^2}
\end{equation}

\begin{figure}[h]
  \begin{center}
  \includegraphics[width=\linewidth]{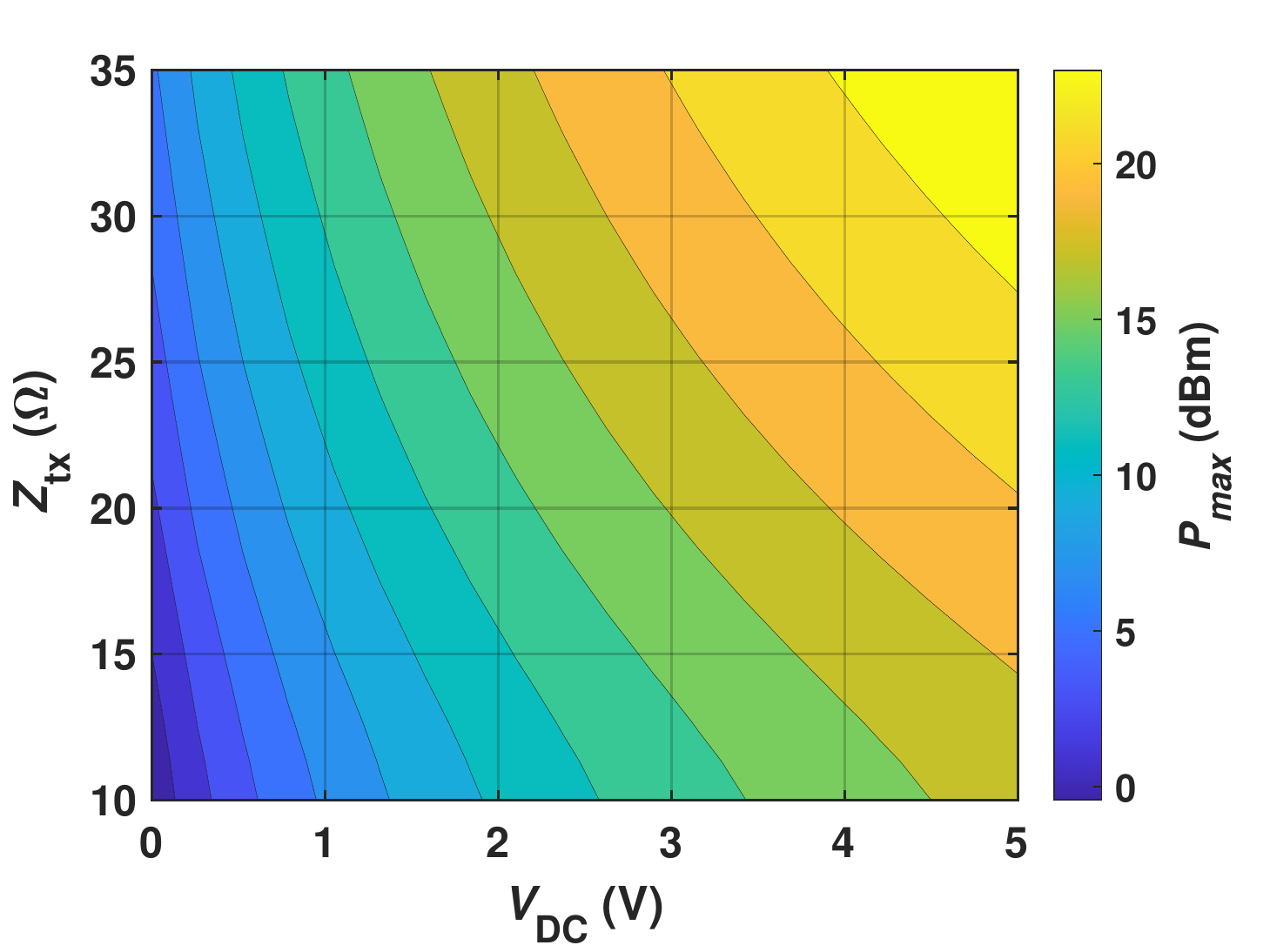}\\
  \caption{ $P_{max}$ value (in dBm) $vs.$ $Z_{tx}$ and $V_{DC}$ for the reflective $pFSL$ in Fig.~\ref{generic_schematic}, computed through \eqref{eq_P_max} after assuming $Z_0$ = 50 $\Omega$, $f_{in}^{opt}$=2.1 GHz, $Q_v$=\QvValue, $V_{bi}$=0.7 V, $C_v$=\CvValue pF and upon validity of the resonance conditions minimizing $P_{th}$.
}\label{Pmax_Vs_VDC_Zth_contour}
  \end{center}
\end{figure}

 Equation \eqref{eq_P_max} aids the selection of the diode by identifying the minimum $V_{DC}$ allowing to preserve a frequency selective limiting behavior for $P_{in}$ values ranging from $P_{th}$ to any desired $P_{max}$ value, given any targeted $C_v$ and $Z_{tx}$ values. As an example, Fig.~\ref{Pmax_Vs_VDC_Zth_contour} shows a contour-plot capturing the analytically derived $P_{max}$ $vs.$ $Z_{tx}$ and $V_{DC}$ [see \eqref{eq_P_max}], assuming the same $\delta$, $f_{in}$ and $Z_0$ values used during the derivation of Fig.~\ref{Cv_Ztx_contour}, a $V_{bi}$ arbitrary set to 0.7 V ($i.e.$, the built-in voltage for silicon diodes) and a $C_v$ value of \CvValue pF [$i.e.$, the same used in our experimental demonstration when considering the additional parasitic capacitance ($\sim$0.4 pF) associated to the diode's package]. As evident from both \eqref{eq_P_max} and Fig.~\ref{Pmax_Vs_VDC_Zth_contour}, $P_{max}$ values lower than 0 dBm and fairly insensitive to $Z_{tx}$ are obtained with low capacitance diodes requiring the use of low $V_{DC}$ values to operate. Nevertheless, significantly higher $P_{max}$ values can still be achieved by selecting larger diodes biased with higher DC voltages. By comparing \eqref{eq_P_max} with \eqref{eq_Pth_Approx}, it can be noticed that $V_{DC}$ represents the most important design parameter to ensure that high $P_{max}$ and low $P_{th}$ values can be achieved simultaneously. Furthermore, contrary to any reported absorptive counterparts, reflective $pFSLs$ are able to protect the integrity of any cascaded electronic components even for $P_{in}>P_{max}$. Yet, within such high-power operative scenario, reflective $pFSLs$ progressively lose their frequency selectivity as $P_{in}$ is increased, thus ultimately showing strong attenuations across significant portions of their original bandwidth.

\section{Designing $pFSLs$ in Commercial Circuit Simulators}

In the previous section, it has been shown how the achievement of the minimum $P_{th}$ and $IL^{s.s}$ values in reflective $pFSLs$ can be ensured for any input frequencies of interest by satisfying four resonant conditions and by minimizing $IL^{s.s}$ [see \eqref{eq_IL_final}]. Therefore, regardless of the nonlinear characteristics exhibited by $pFSLs$, the synthesis of the passive impedances [i.e., [$Z_T$], $Z_a$,  $Z_b$  and $Z_c$  (see Fig.~\ref{generic_schematic})] forming any board-level reflective $pFSLs$ can be accomplished through linear simulation and optimization techniques, after selecting an available diode with the lowest possible $C_v$ and the highest possible $Q_v$, given the minimum tolerated $P_{max}$ value. It is worth pointing out that the capability to synthesize the different components of $pFSLs$ through linear methods enables the reliance on conventional algorithms during the optimization of both the $pFSLs$’ circuit and layout. Consequently, the optimal design conditions for any $pFSLs$, including those operating at high frequency and consequently more impacted by layout parasitics, can be identified more reliably and more easily than what is possible when only relying on perturbation-based techniques. Nevertheless, such techniques are still required to assess the behavior of $pFSLs$ for $P_{in}>P_{th}$, thus in the operative regime where the evolution of the circuit parameters strongly depends on the nonlinearities of the diode. Among the existing perturbation-based techniques, the power auxiliary generator (pAG\cite{Cassella2015}) technique provides unique means to extract the steady-state response of any parametric circuits without having to numerically enforce the validity of the non-perturbation condition\cite{Suarez2009} due to the adoption of an auxiliary generator in the circuit. A pAG consists of a continuous-wave (CW) voltage generator operating at $f_d$ and delivering a small non-perturbative power through a generator impedance ($Z_G$). When a pAG is used to simulate the response of a $pFSL$ through a commercial Harmonic Balance (HB) circuit simulator, the pFSL’s output termination (i.e., $Z_0$ in the circuit in Fig.~\ref{generic_schematic}) must be replaced by a pAG with $Z_G$ also equal to $Z_0$. This allows to insert $f_d$ in the vector of frequencies that HB processors use to evaluate the response of driven RF circuits. In addition, by ensuring that $Z_G$ matches $Z_0$, the adoption of the pAG does not cause variations of the impedance seen by the diode at $f_d$ or at multiples of $f_d$, thus preventing any undesired changes in the dynamics of the circuit which would lead to unreliable predictions of the circuit's response. The extraction of the steady-state response of $pFSLs$ can be achieved by sweeping $P_{in}$ from a much lower power than $P_{th}$ up to the maximum power level of interest, while configuring the values of the computed current and voltage phasors for any evaluated power level as initial conditions for the same circuit parameters prior the computation of the following data point to assess. Through this $ad$-$hoc$ sweeping strategy, it is possible to extract $IS$ by evaluating the trend of the $pFSLs$’ insertion-loss for power levels exceeding $P_{th}$. This also provides the means to fine-tune some of the component values synthesized in the earlier design stage towards the achievement of the highest $IS$ value. In order to do so, it is convenient to look at $Z_{in}^{\omega_{in}}$  (see Fig.~\ref{generic_schematic}) while varying some strategically selected circuit components to let the magnitude of this impedance be as small as possible for $P_{in}$ exceeding $P_{th}$. During this design phase, the components synthesizing $Z_b$ are the most adequate to be fine-tuned as they do not affect $IL^{s.s}$ while only slightly altering $P_{th}$. It is also important to point out that the ability to accurately predict $IS$ in any $pFSLs$ is heavily influenced by the reliability of the available nonlinear circuit model for the adopted diode.

\begin{figure}[h]
\begin{subfigure}{1.0\linewidth}
    \centering
    \caption{}
    \includegraphics[width=0.7\linewidth]{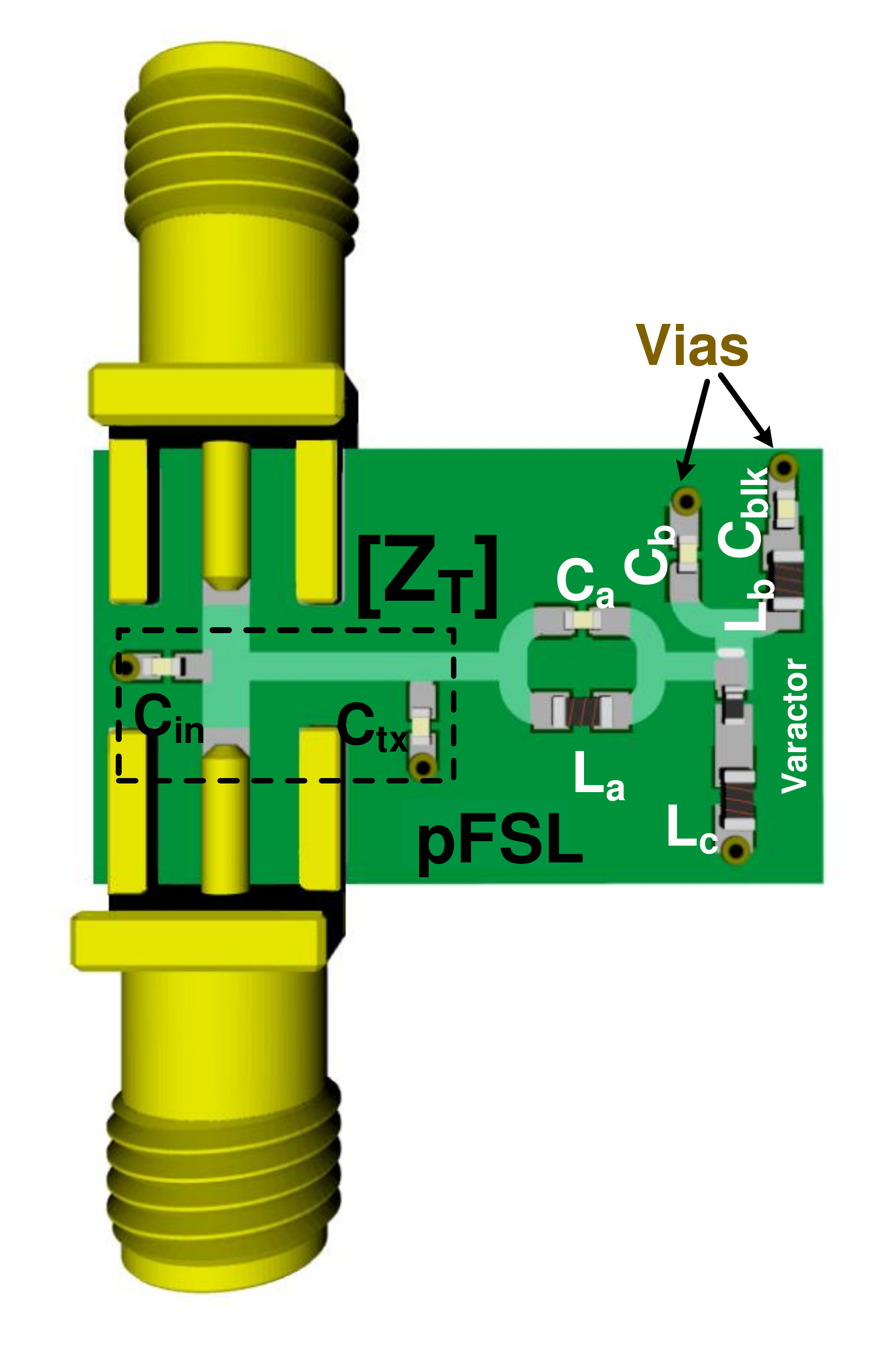} 
    \label{3D_FSL}
\end{subfigure}
\begin{subfigure}{1.0\linewidth}
    \centering
    \caption{}
    \includegraphics[width=0.6\linewidth]{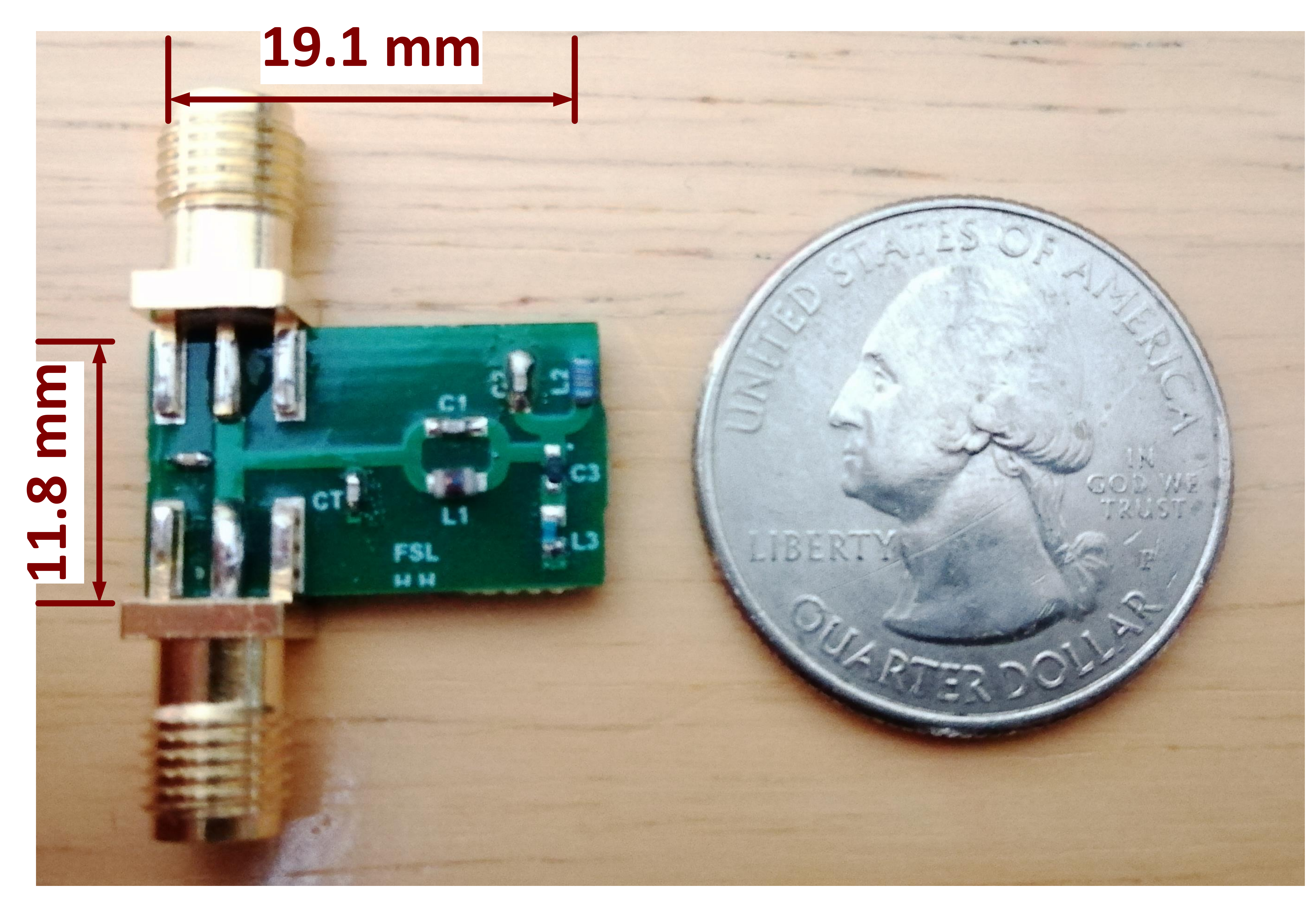} 
    \label{PCB_Picture}
\end{subfigure}
\caption{ a) 3D-view of the designed $pFSL$ showing [$Z_T$] and the area dedicated to each adopted lumped component; b) top-view photo of the built $pFSL$ circuit. The model numbers for the adopted lumped components are: i) 0402DC-11NXGRW for $L_a$; ii) 0603HP-6N8XGLU for $L_b$; iii) GJM1555C1H1R4WB01D for $C_a$; iv) GJM1555C1HR30WB01D for $C_b$; v) 0402DC-1N0XJRW for $L_c$; vi) \VaractorModelNumber~ for the diode; vii) GJM1555C1H1R5WB01D for $C_{tx}$; viii) GJM1555C1H1R0WB01D for $C_{in}$, and ix) GJM1555C1H120FB01 for $C_{blk}$.}
\label{PCB_3D}
\end{figure}

\section{Experimental Results}

In order to demonstrate the capability of reflective $pFSLs$ to simultaneously exhibit low $IL^{s.s}$, low $P_{th}$ and high $IS$ values, we designed a FR-4 Printed Circuit Board (PCB) implementation of a $\sim$\FinOperatingApproxGHz GHz reflective $pFSL$ (i.e., $f_{in}^{opt}\sim$\FinOperatingApproxGHz GHz) using a commercial off-the-shelf hyperabrupt varactor (\VaractorModelNumber) as a diode. The assembled $pFSL$ relies on a hybrid implementation of the circuit topology shown in Fig.~\ref{generic_schematic}, including a set of lumped components strategically selected together with the optimal geometrical layout  characteristics ($i.e.$, the 
 \begin{figure}[h]
  \begin{center}
  \includegraphics[width=\linewidth]{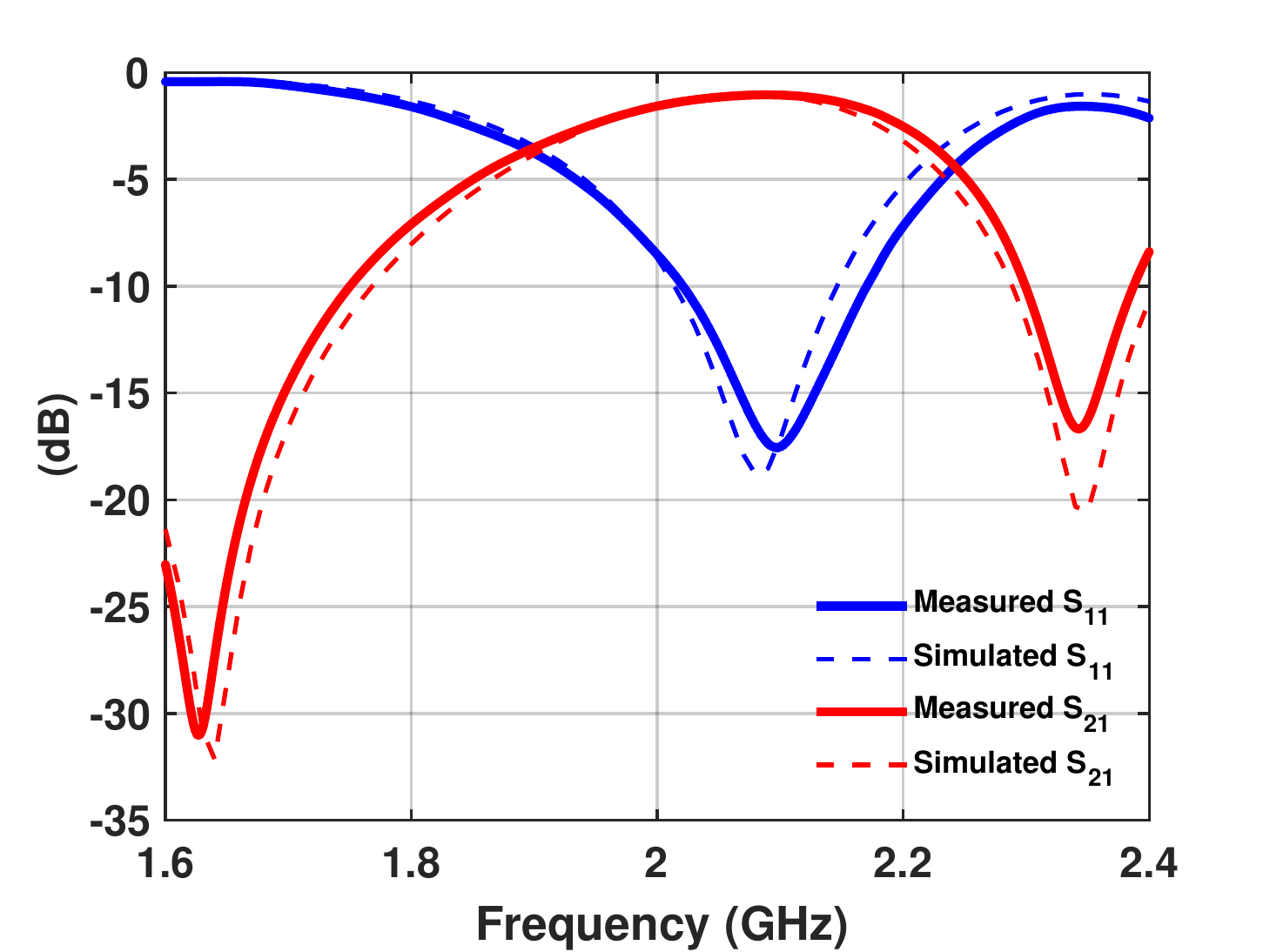}\\
  \caption{ Measured (continuous lines) and simulated (dotted lines) $S_{11}$ and $S_{21}$ of the reported $pFSL$ for frequencies included within the $pFSL$'s bandwidth and when considering an input power level (-30 dBm) during the frequency sweep much lower than the expected $P_{th}$.
}\label{S11_S21_Vs_Freq_MeasVsSim}
  \end{center}
\end{figure}
length and width of all interconnecting lines, the shape and foot-print of the selected surface-mount connectors, and the radius of each circular via). This allows to achieve a small form factor (\PCBAreacm $cm^2$ of board area) and the lowest possible $P_{th}$ given an $IL^{s.s}$ lower than a maximum tolerated value (here set to 1 dB). In particular, $Z_a$ and $Z_b$  were synthesized through two $LC$ parallel resonators with inductors $L_a$ and $L_b$, and capacitors $C_a$ and $C_b$; whereas $Z_c$  was synthesized through an inductor ($L_c$) in series with the selected varactor. Also, a DC-blocking capacitor ($C_{blk}$ = 12 pF) was introduced in series with $L_b$ to permit the DC-biasing of the selected diode through a Bias-Tee (\BiasTeeModelNumber) at the input port, even allowing an analog reconfigurability of $P_{th}$ and $f_{in}^{opt}$, as discussed later. The transformation stage at $f_{in}^{opt}$ was designed to exhibit a $Z_{tx}$ value close to \ZtxValue $\Omega$ when relying on two lumped capacitors ($C_{tx}$ and $C_{in}$) and two distributed inductors made of short lines. Such a $Z_{tx}$ value was chosen to ensure $P_{th}$ and $IL^{s.s}$ values lower than -3 dBm and 1 dB respectively (see Fig.~\ref{Cv_Ztx_contour}) based on the $C_v$ (\CvValue pF) and $Q_v$ ($\approx$\QvValue) values exhibited by the varactor when $V_{DC}$ is set to \Vbias V. Also, the adoption of distributed components in the transformation stage represents the most convenient design solution to ensure that the performance is not degraded by the capacitive coupling between the connectors’ foot-prints and the rest of the circuit, while preserving the lowest possible form-factor. The designed $pFSL$ is visualized in Fig.~\ref{PCB_3D} together with a photo of its built implementation, where the values and the model numbers of the selected lumped components are listed in the caption. 
The measured electrical response of the reported $pFSL$ was first characterized through the extraction of its S-parameters for driving power levels (e.g., -30 dBm) much lower than the expected limiting threshold, thus in the operative regime where a linear operation can be assumed. Fig.~\ref{S11_S21_Vs_Freq_MeasVsSim} shows the resulting measured and closely matching simulated plots of the $pFSL$'s $S_{21}$ and $S_{11}$ $vs.$ $f_{in}$. Evidently, the reported $pFSL$ exhibits a band-pass characteristic with a measured 3-dB fractional bandwidth (BW) of \FBW $\%$. Also, a minimum $IL^{s.s}$ of \ILMin dB was measured for a frequency (\FinOperatingGHz GHz) close to the targeted $f_{in}^{opt}$. Such a record-low $IL^{s.s}$ value closely matches both the corresponding circuit simulated one  (\ILSimulation dB) and our analytically derived expectation (\ILAnalytical dB, see Fig.~$\ref{IL_Vs_Cv_Ztx_contour}$).


\begin{figure}[h]
    \centering
\def\big{\includegraphics[width=1.0\linewidth]{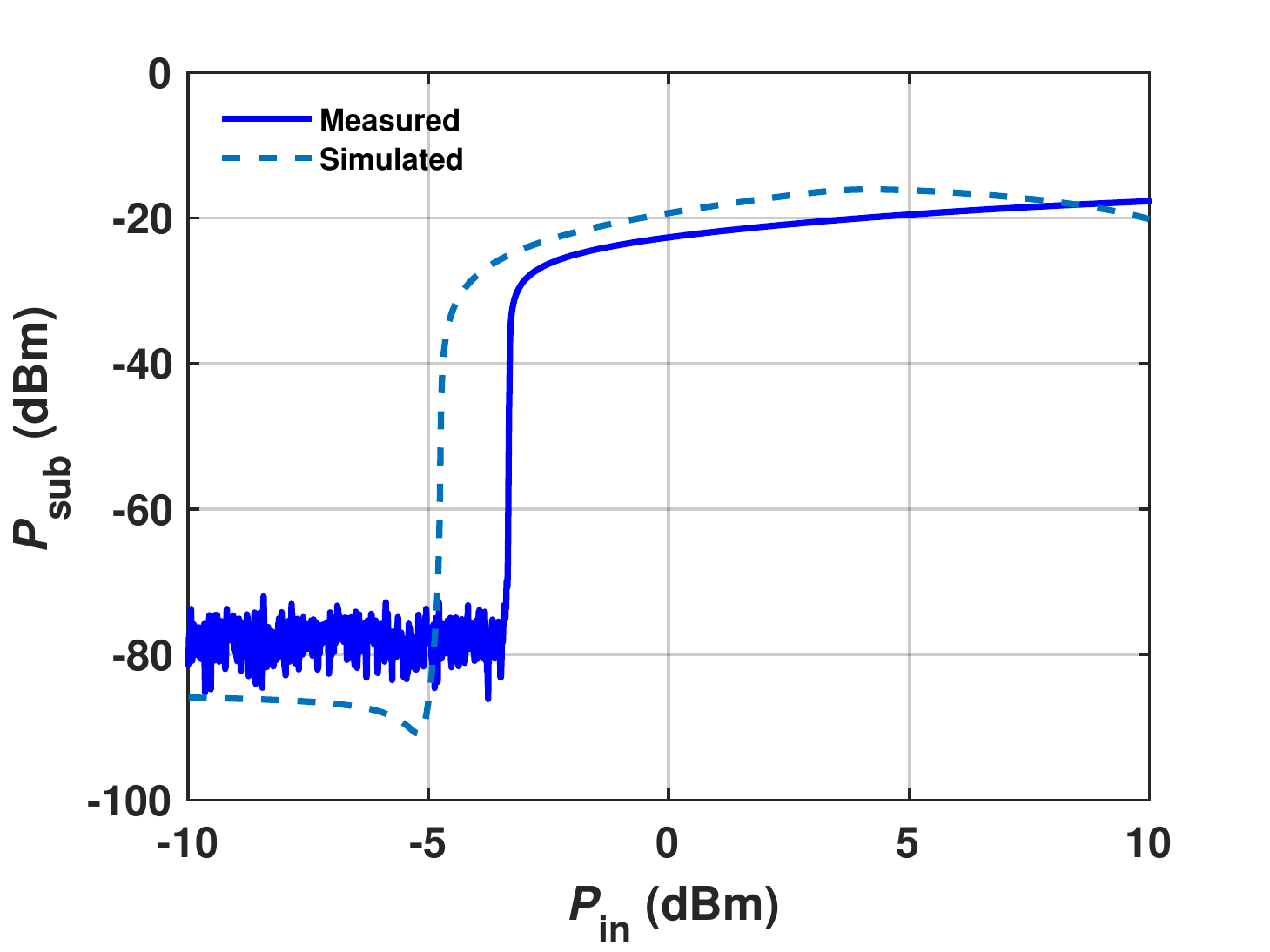}}  
\def\little{\includegraphics[width=1.6in]{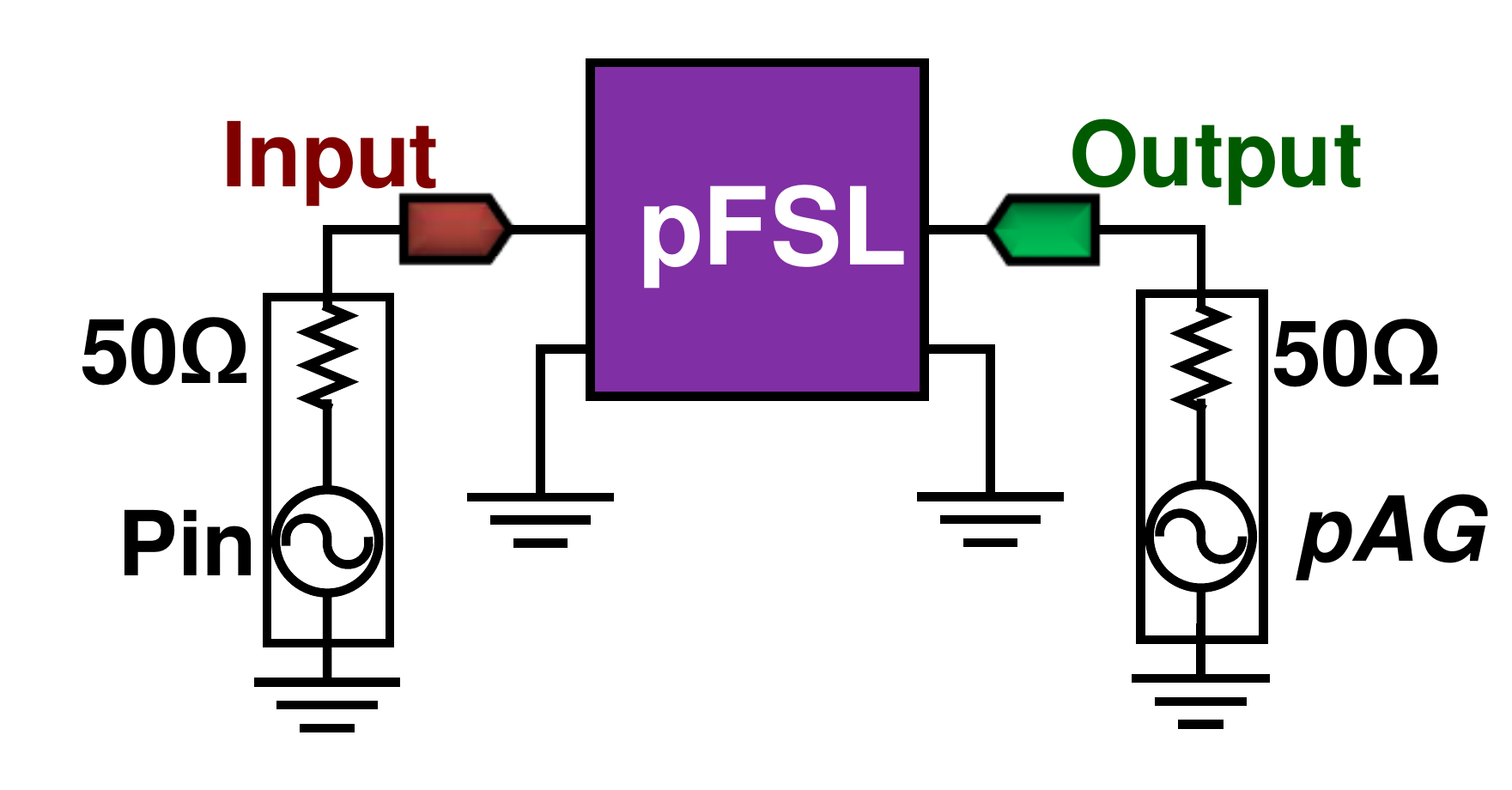}}
\def\stackalignment{r}  
\topinset{\little}{\big}{78pt}{28pt}   
\caption{Measured (in blue) and simulated (in red) trends of $P_{sub}$ $vs.$ $P_{in}$. A simplified schematic is also shown in the inset, describing how we used a pAG erogating a non-perturbative power (-90 dBm) at 1.03 GHz in order to extract the simulated data. }
\label{Bifurcation_MeasVsSim}
\end{figure}

\begin{figure}[h]
\begin{subfigure}{1.0\linewidth}
    \centering
    \caption{}
    \def\big{\includegraphics[width=1.0\linewidth]{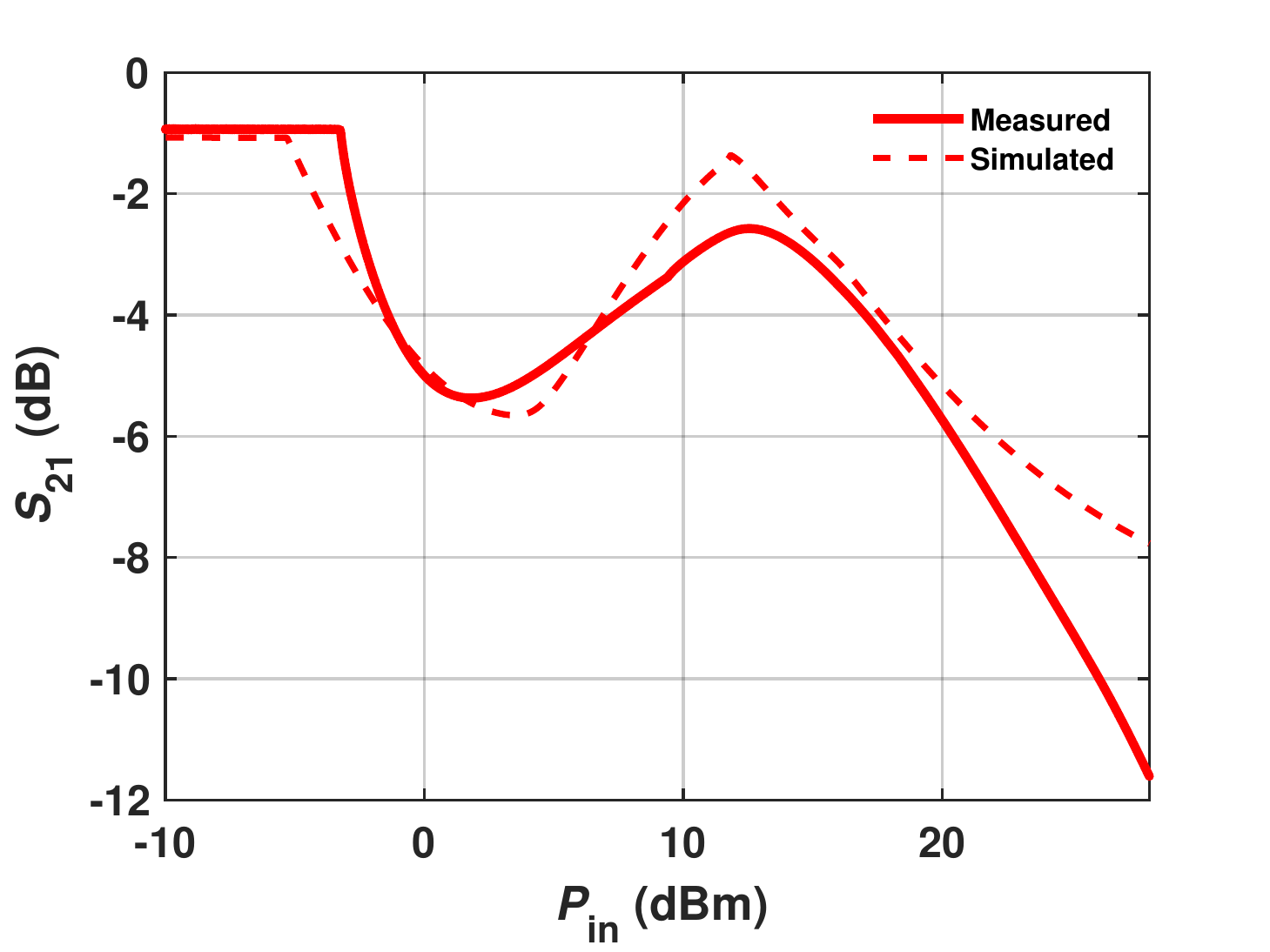}}  
    \def\little{\includegraphics[width=1.39in]{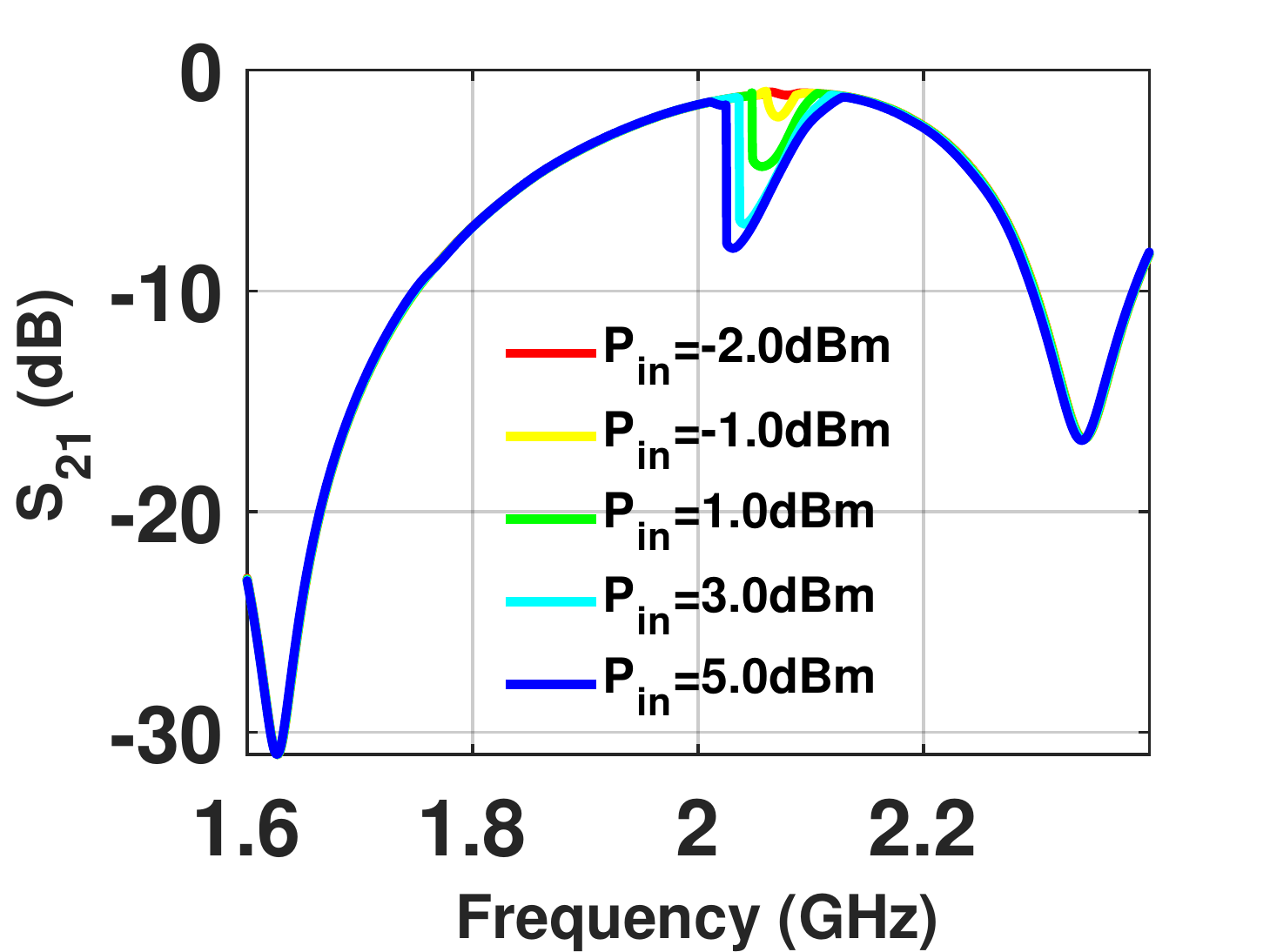}}
    \def\stackalignment{l}  
    \topinset{\little}{\big}{83pt}{34pt}   
    \label{S21_Vs_Power_MeasVsSim}
\end{subfigure}
\begin{subfigure}{1.0\linewidth}
    \centering
    \caption{}
    \includegraphics[width=1.0\linewidth]{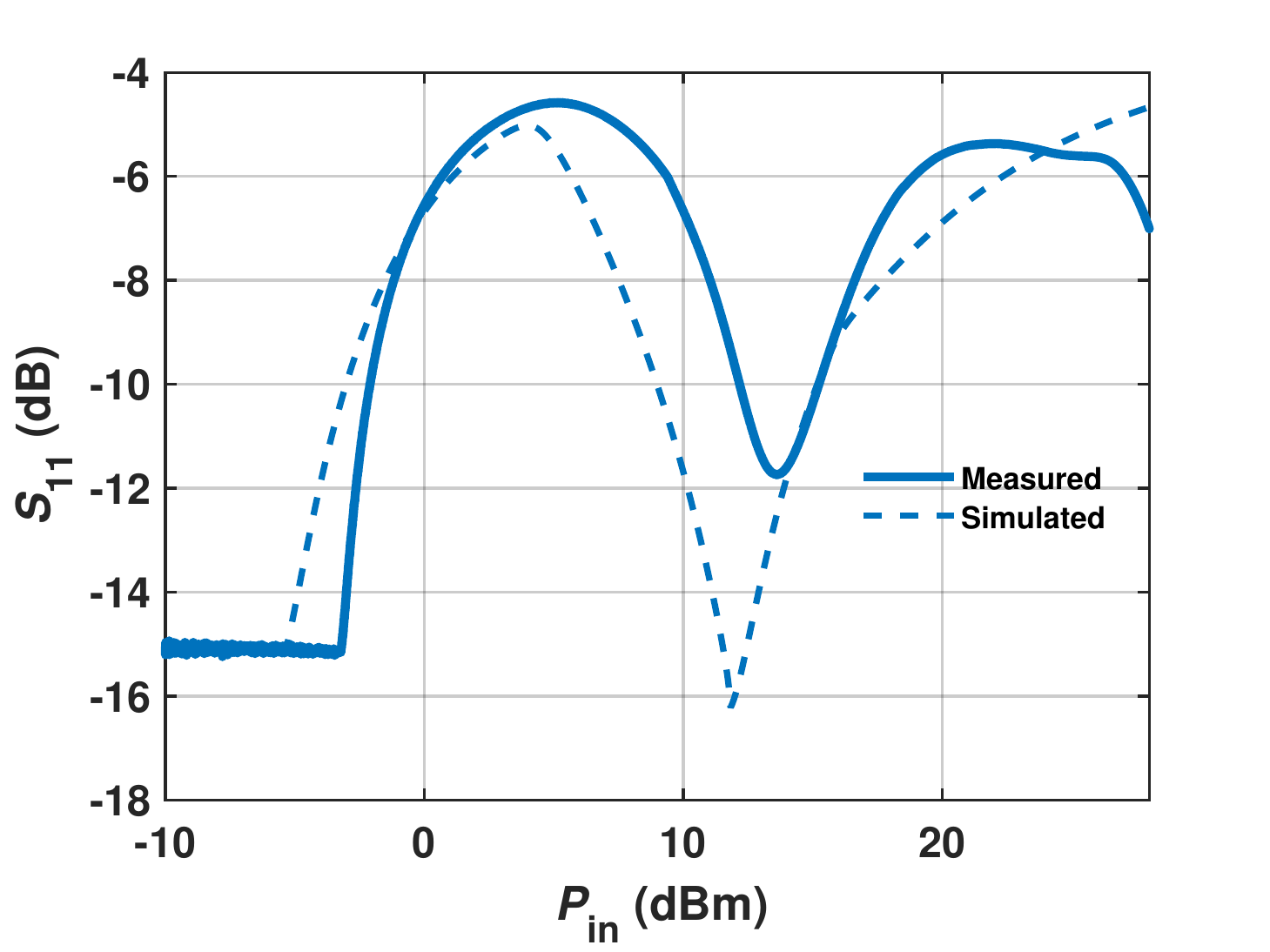} \label{S11_Vs_Power_MeasVsSim}
\end{subfigure}
\caption{ Measured (continuous lines) and simulated (dotted lines) $S_{21}$ (a) and $S_{11}$ (b) at \FinOperatingGHz GHz for $P_{in}$ values ranging from \PinSweepMin dBm to \PinSweepMax dBm. The measured trend of the $S_{21}$ $vs.$ $f_{in}$ for different input power levels used during the sweep, ranging from $P_{th}$ to $P_{max}$, is also shown in the inset of (a) to highlight the frequency selectivity of the parametrically triggered limiting mechanism exploited by the reported $pFSL$ for $P_{in}<P_{max}$.}
\label{S_Vs_Power_MeasVsSim}
\end{figure}

 Later, we characterized the $pFSL$ response for $P_{in}$ approaching and exceeding $P_{th}$. In order to most reliably extract $P_{th}$, we found the minimum $P_{in}$ value around \FinOperatingApproxGHz GHz triggering a 2:1 sub-harmonic oscillation in the circuit. In order to do so, we identified the power level at which a bifurcation triggering a 2:1 frequency division occurs from the measured output power ($P_{sub}$, see Fig.~\ref{Bifurcation_MeasVsSim}) at half of any explored driving frequencies. It is worth emphasizing that $P_{sub}$ is not trivial only for $P_{in}$ exceeding $P_{th}$. As shown in Fig.~\ref{Bifurcation_MeasVsSim}, the measured $pFSL$ exhibits a minimum $P_{th}$ of \PthMin dBm at \FinOperatingGHz GHz ($i.e.$, the same frequency minimizing $IL^{s.s}$, see Fig.~\ref{S11_S21_Vs_Freq_MeasVsSim}). Such a measured $P_{th}$ value matches closely the predicted one found from the circuit simulated trend of $P_{sub}$ $vs.$ $P_{in}$ (see the dotted line in Fig.~\ref{Bifurcation_MeasVsSim}). This simulated trend was obtained by utilizing the pAG technique and the extracted electromagnetic model (EM) of the designed board, together with the S-parameters of the selected lumped components. The measured $P_{th}$ is also close to its analytically predicted value (\PthAnalytical dBm, see Fig.~\ref{Pth_Vs_Cv_Ztx_contour}), given the capacitance of the selected diode and the value chosen for $Z_{tx}$. 
The $IS$ value of the built $pFSL$ at \FinOperatingGHz GHz was assessed as well. This was done by extracting the corresponding $S_{21}$ for $P_{in}$ values ranging from \PinSweepMin dBm to \PinSweepMax dBm (the maximum available power level in our experiment,  Fig.~\ref{S21_Vs_Power_MeasVsSim}). As evident, significant $IS$ values up to \ISMaxLessThanPmax dB were found for $P_{in}>P_{th}$ and lower than \PMaxdBm dBm. Within this power range, the designed $pFSL$ shows a trend of the large-signal $S_{21}$ $vs.$ $f_{in}$ that realizes a frequency selective notch centered around \FinOperatingGHz GHz, clearly indicating the activation of a frequency selective attenuation as $P_{in}$ is increased above $P_{th}$ (see the inset of Fig.~\ref{S21_Vs_Power_MeasVsSim}). Furthermore, despite its less frequency selective limiting behavior for $P_{in}>P_{max}$, the measured $pFSL$ shows high $IS$ values even for $P_{in}$ higher than \PMaxdBm dBm ($IS>$\ISMaxAtThirthydBm dB for $P_{in}$ approaching \PinSweepMax dBm). It is worth emphasizing that the achievement of such high $IS$ value is granted by a parametrically triggered nonlinear mechanism causing the return-loss at the two $pFSL$'s ports to significantly increase as $P_{in}$ is made larger than $P_{th}$ (see Fig.~\ref{S11_Vs_Power_MeasVsSim}). 

 \begin{figure}[h]
  \begin{center}
\begin{subfigure}{1.0\linewidth}
    \centering
    \caption{}
    \includegraphics[width=1.0\linewidth]{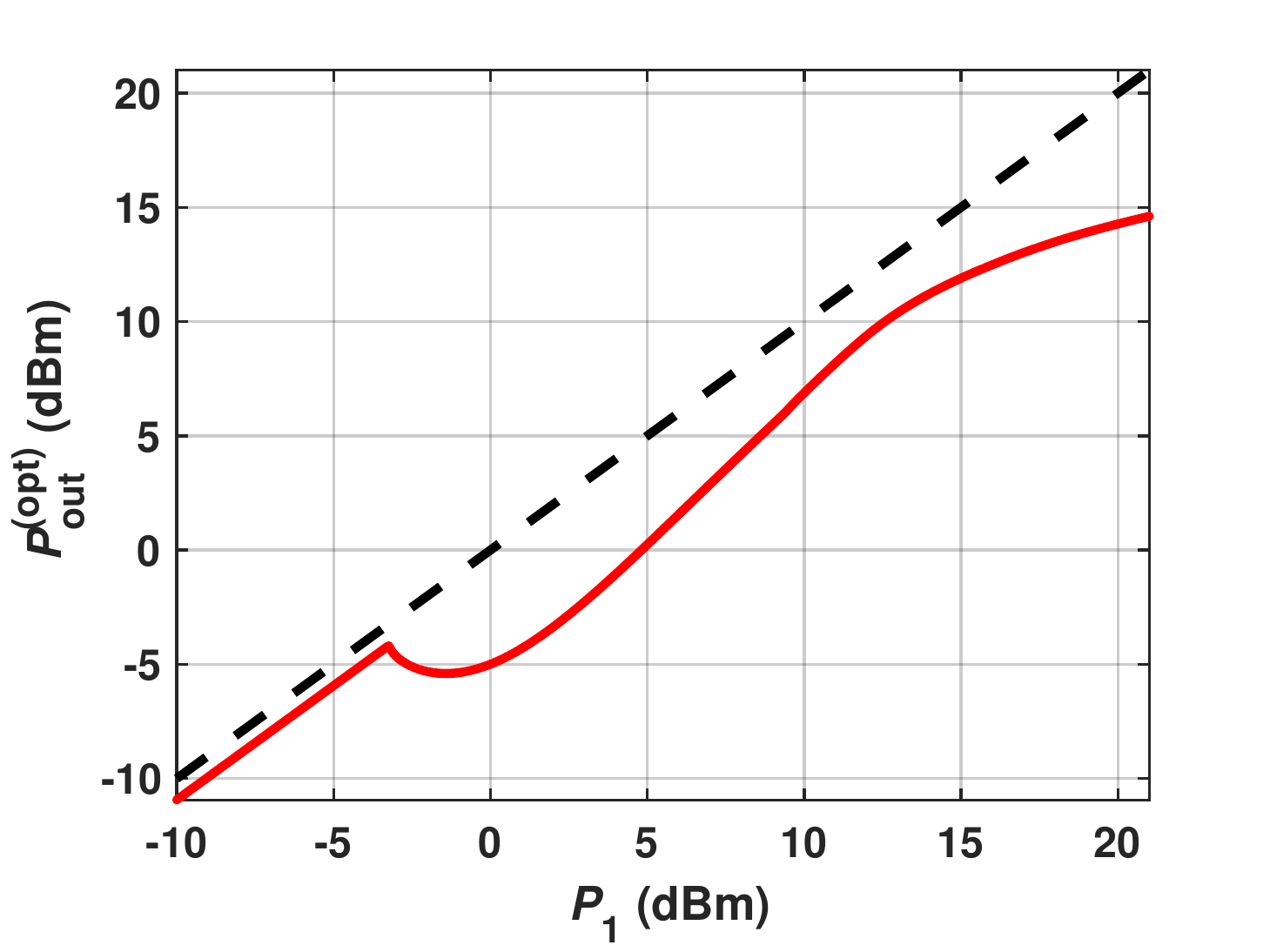} 
    \label{Pout_Vs_Pin_Meas}
\end{subfigure}
\begin{subfigure}{1.0\linewidth}
    \centering
    \caption{}
    \includegraphics[width=1.0\linewidth]{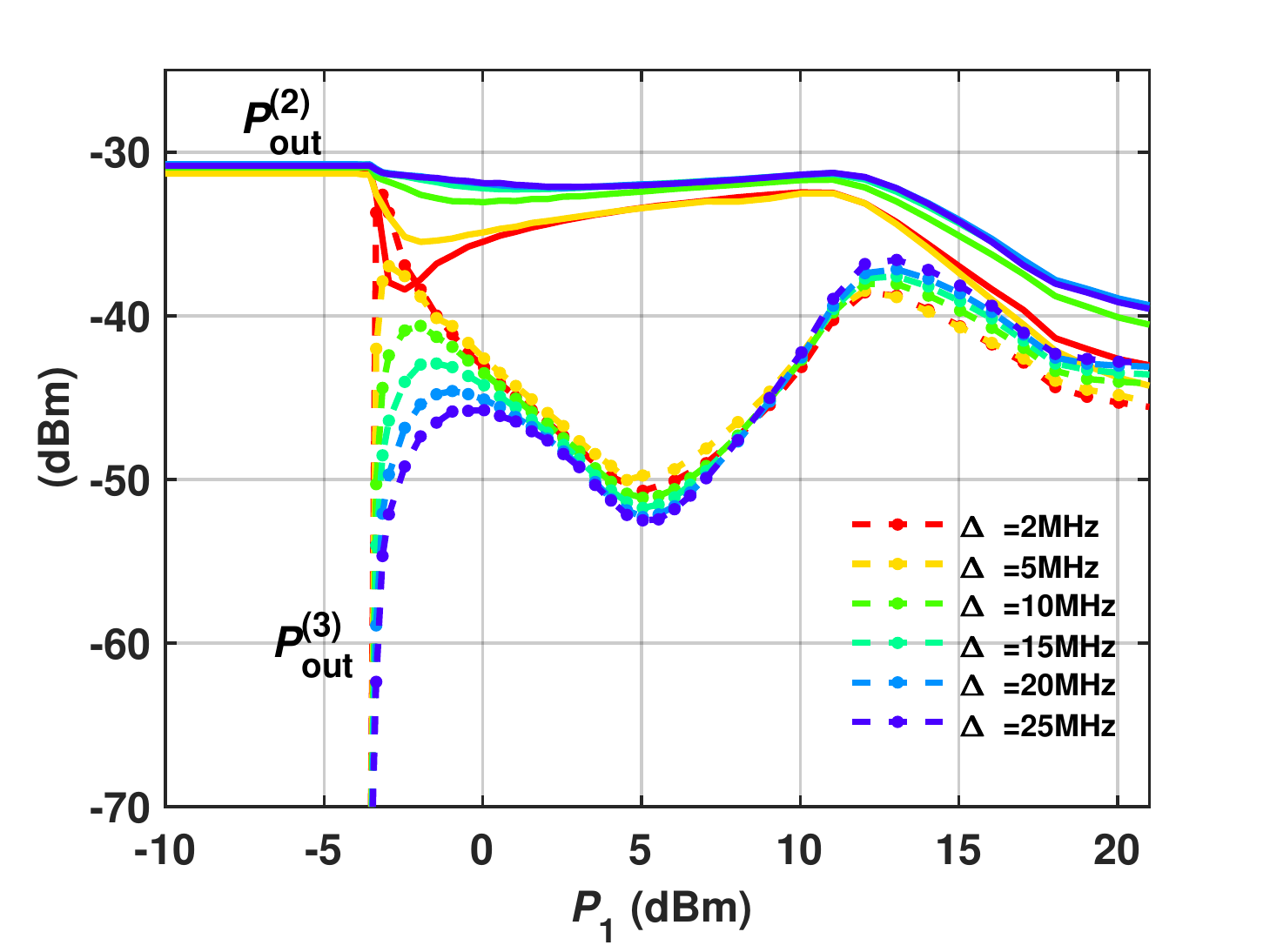} 
    \label{PIM_Psig_Vs_Pin_Meas}
\end{subfigure}
\caption{ a) A comparison of the measured trend of $P_{out}^{opt}$ $vs.$ $P_{1}$ (solid red line) with the corresponding trend expected from a linear and loss-less two-port networks (black dashed line); b) measured trends of $P_{out}^{(2)}$ (solid lines) and $P_{out}^{(3)}$ (dashed lines) $vs.$ $P_1$ for $\Delta$ values ranging from 2 MHz to 25 MHz.
}\label{}
  \end{center}
\end{figure}

 \begin{figure*}[t]
\centering
    \begin{subfigure}[b]{0.32\linewidth}
    \centering
    \caption{}
    \includegraphics[width=1.0\linewidth]{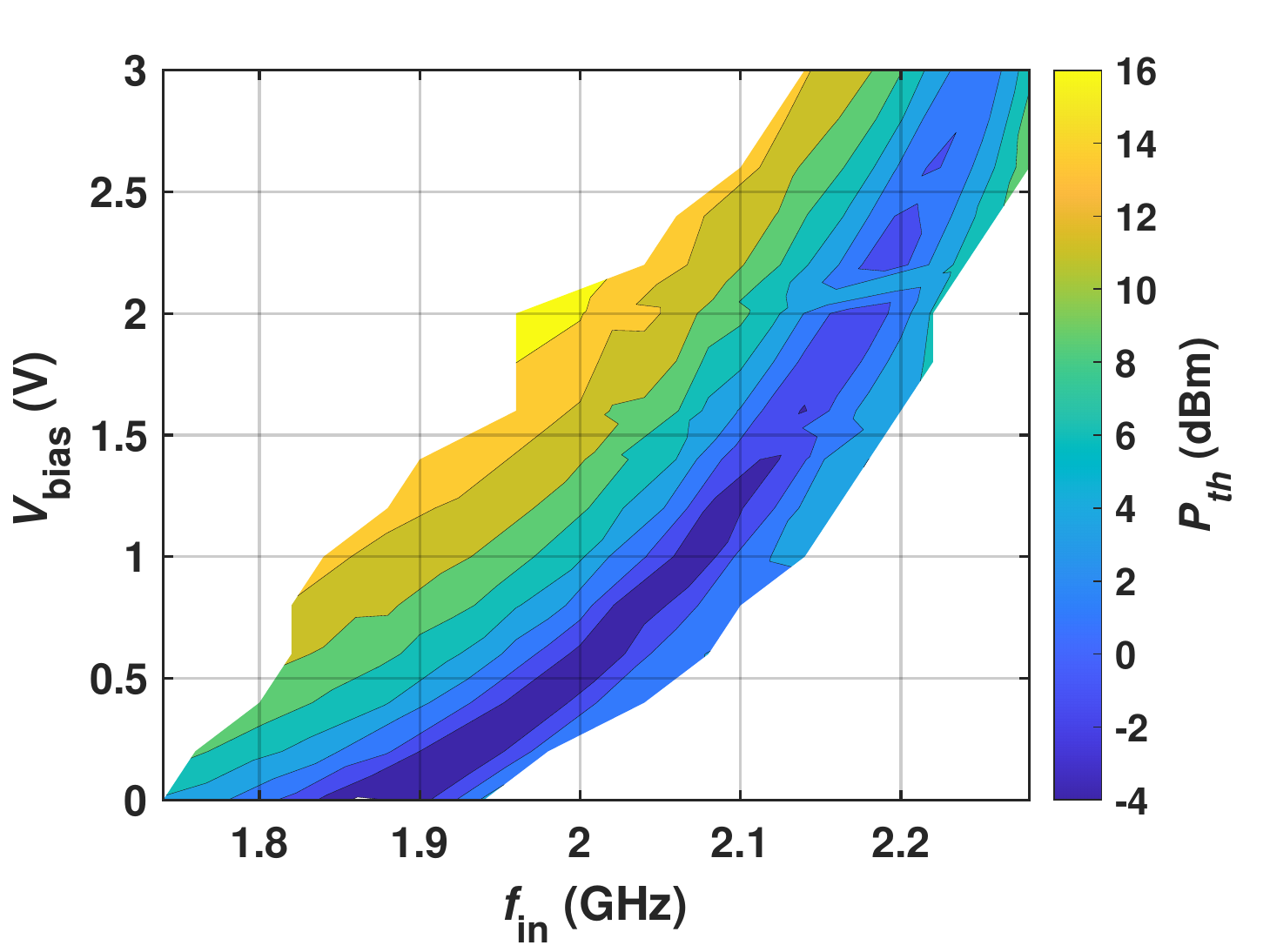} \label{contour_Pth}
\end{subfigure}
\begin{subfigure}[b]{0.32\linewidth}
    \centering
    \caption{}
    \includegraphics[width=1.0\linewidth]{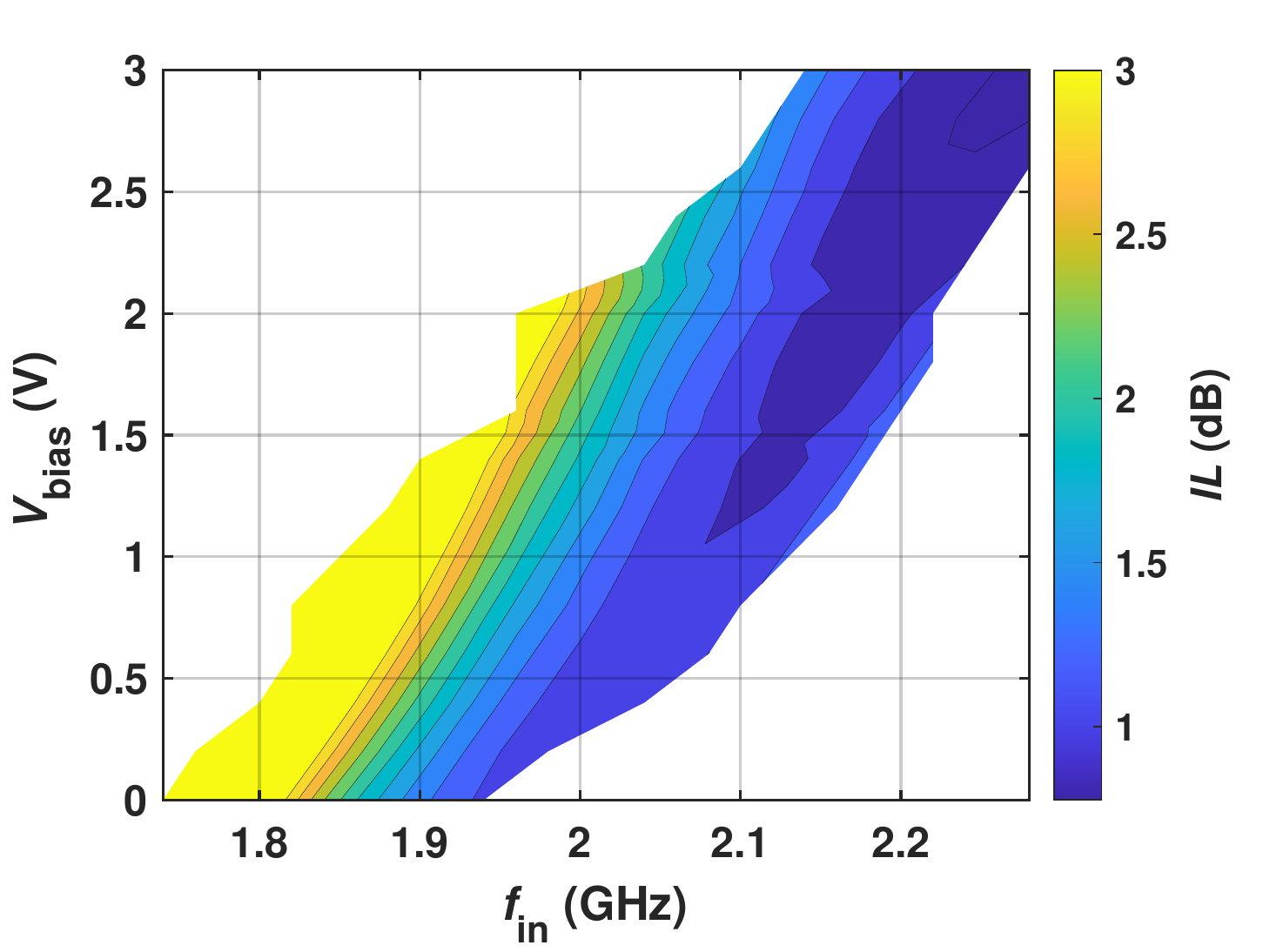} 
    \label{contour_IL}
\end{subfigure}
\begin{subfigure}[b]{0.32\linewidth}
    \centering
    \caption{}
    \includegraphics[width=1.0\linewidth]{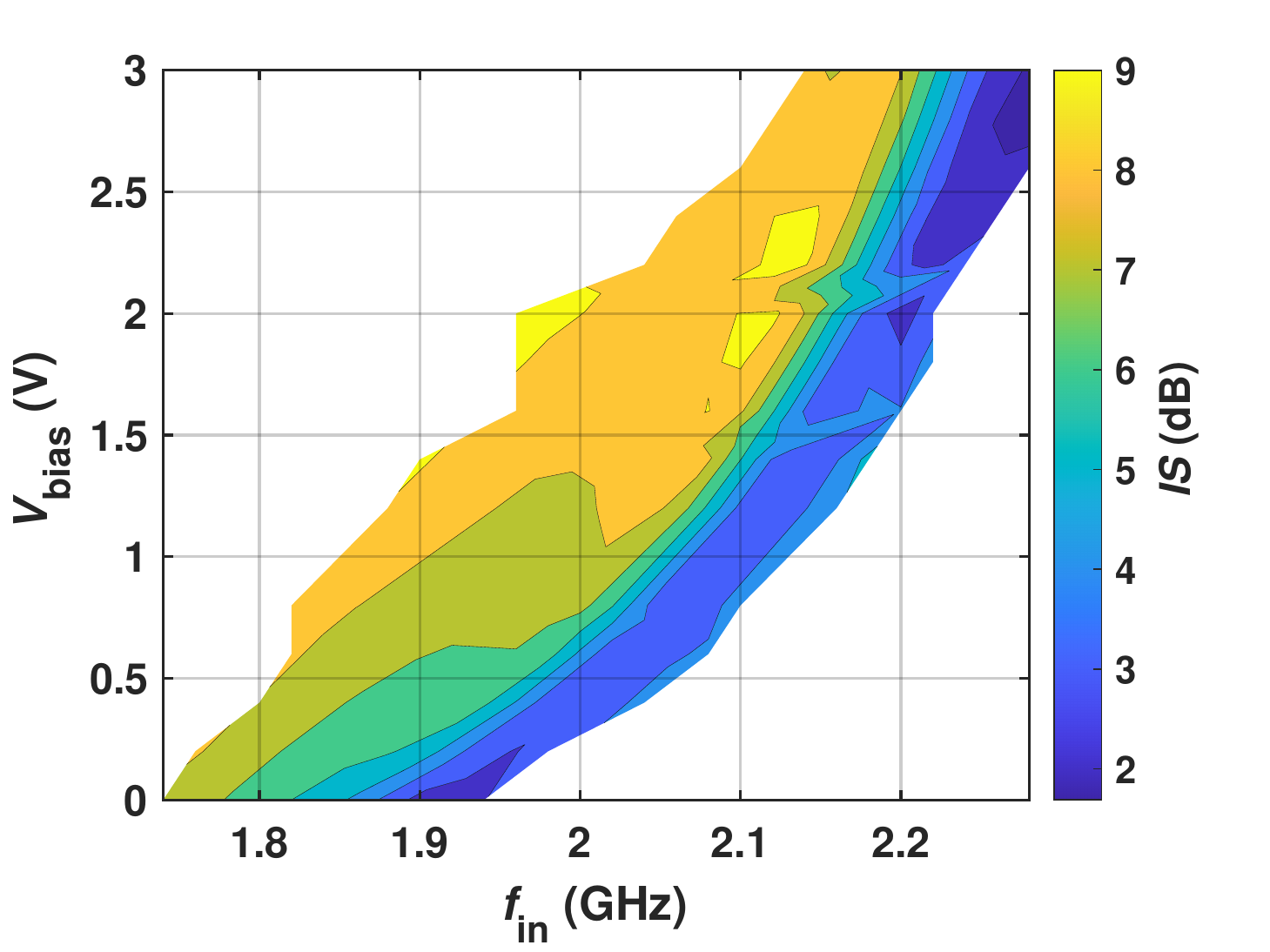} 
    \label{contour_IS}
\end{subfigure}
\caption{ Contour-plots capturing the measured $P_{th}$ (a), $IL^{s.s}$ (b) and $IS_{max}^{<P_{max}}$ (c) values $vs.$ $V_{DC}$ and $f_{in}$.}
\label{Measured_Contour_Plots}
\end{figure*}

Nevertheless, both the measured $S_{21}$ and $S_{11}$ trends $vs.$ $P_{in}$ (Fig.~\ref{S_Vs_Power_MeasVsSim}) do not show a fully monotonic behavior, as also expected from the corresponding simulated trends (see the dotted lines in Fig.~\ref{S_Vs_Power_MeasVsSim}). This is due to changes in the diode's dynamics becoming more and more significant as the input power approaches $P_{max}$. In particular, through our circuit simulations, we found that the $P_{in}$ value associated with the local maximum (minimum) of the $S_{21}$ ($S_{11}$) for $P_{in}>P_{th}$ corresponds to a peak voltage level across the adopted diode equal to $V_{bi}+V_{DC}$. This finding provided us with simple means to identify the $P_{max}$ value [\PMaxdBm dBm, close to its estimated analytical value (\PMaxAnalyticaldBm dBm, see Fig.~\ref{Pmax_Vs_VDC_Zth_contour})] of the built $pFSL$ by extracting the $P_{in}$ level at which the same phenomenological change occurs in both the measured $S_{21}$ and $S_{11}$ trends. Finally, we evaluated the performance of the built $pFSL$ when simultaneously driven by a \FinOperatingGHz GHz signal and by a much lower power (\PinLowPowerSigdBm dBm) in-band tone. In order to do so, we relied on a conventional $two$-$tone$ harmonic test where two RF signals, combined through an external power combiner (\CombinerModelNumber), were simultaneously injected in the circuit from the $pFSL$'s input port. In particular, the signal at \FinOperatingGHz GHz, with power labeled as $P_1$, emulated the presence of a strong EMI at the $pFSL$'s input port whereas the second much lower power signal (with -30 dBm), detuned from \FinOperatingGHz GHz by an amount labeled as $\Delta$ ($i.e.$, $f_2$= \FinOperatingGHz GHz+$\Delta$), was used to emulate the presence of a simultaneously received low-power signal carrying useful information. The output power levels at \FinOperatingGHz GHz  ($P_{out}^{opt}$), at $f_2$ ($P_{out}^{(2)}$) and the one ($P_{out}^{(3)}$) at the strongest $3^{rd}$-order intermodulation product ($f_3$ = 4.12GHz-$f_2$) were measured for $P_{1}$ levels ranging from \PinSweepMin dBm to 21 dBm. It is crucial to emphasize that achieving low $P_{out}^{(3)}$ values is particularly important in applications where contiguous in-band channels with small frequency separations can be simultaneously received. The measured trends of $P_{out}^{opt}$, $P_{out}^{(2)}$ and $P_{out}^{(3)}$ $vs.$ $P_1$ are reported in  Fig.~\ref{Pout_Vs_Pin_Meas}  and Fig.~\ref{PIM_Psig_Vs_Pin_Meas} for $\Delta$ values varying between 2 MHz and 25 MHz. It can be seen that the parametric mechanism responsible for the suppression of $P_{out}^{opt}$ also causes an undesired attenuation ($\alpha$) of $P_{out}^{(2)}$ that is higher for small values of $\Delta$ (see Fig~\ref{PIM_Psig_Vs_Pin_Meas}). Yet, $\alpha$ values not exceeding 2 dB and 4 dB were attained for $\Delta$ values of 10 MHz and 5 MHz, thus demonstrating a good frequency selectivity in the limiting operation generated by the circuit. Also, $P_{out}^{(3)}$ values below -40 dBm were measured for $P_{in}$ values lower than $P_{max}$ and for $\Delta$ higher than 10 MHz. It is key to point out that the ability to achieve such low $P_{out}^{(3)}$ values, regardless of the active parametrically triggered power limiting behavior, is due to the low impact exerted by the varactor's quadratic nonlinearities on the circuit dynamics for power levels that are lower than $P_{max}$. Nevertheless, for $P_{in}$ larger than $P_{max}$ the compressing and non-frequency-selective diode's electrical response determines a saturation in the power level at all frequencies in the circuit, causing the observed behavior for $P_1>$ \PMaxdBm dBm. Ultimately, while the observation of $P_{sub}$ has provided us with reliable means to quantify $P_{th}$, the corresponding parametrically generated sub-harmonic signal, even if small, can also slightly degrade the signal-to-noise ratio ($SNR$) at the $pFSL$’s output. Nevertheless, due to the large frequency separation between the $pFSL$’s main operational frequency and its sub-harmonic one, a strong attenuation of $P_{sub}$ can still be achieved  through the adoption of a proper filtering stage at the $pFSL$ output, without altering the circuit dynamics.


\begin{table*}[t]
\centering
\caption{A performance comparison between the reported $pFSL$ and other previously reported diode-based prototypes. Note: the reported $IS^{28dBm}$ values marked with a "*" were estimated through a linear interpolation of available measured data that refer to a maximum $P_{in}$ value lower than \PinSweepMax dBm.}
\begin{tabular}{|c| c | c | c | c | c | c | c |} 
 \hline
                            & Tech.              & $P_{th}$      & $IL^{s.s.}$   & $f_{in}$      & $IS_{max}^{<P_{max}}$      &$IS^{28dBm}$    & Component Area  \\ 
                            &                   & $(dBm)$      & $(dB)$         & $(MHz)$       & $(dB)$                        & $(dB)$        & $(cm^{2})$        \\ 
\hline
\cite{Ramirez2008}          &  Parametric      &    4.5        &     3          &    850         &  9                       &  $<$ 3 dB*    & n/a \\ 
 \hline
\cite{Wolf1960}             &  Parametric      &    10        &     3          &    2200         &  11                      & 20            & n/a  \\ 
 \hline
\cite{Ho1961}               &  Parametric      &    3        &     3          &    2380         &  13                       & $<$ 18 dB*     & $\sim$9 \\ 
 \hline
\cite{Phudpong2009}         &  BSF              &    3        &     2          &    1000         &  13                      &  n/a           &  $\sim$37 \\ 
 \hline
\cite{Naglich2016}          &  BPF              &    24        &     2          &    1500         &  10                   &     10          & 378 \\ 
 \hline
\cite{Hueltes2017}          &  Coupler          &    6        &     1.5          &    2000         &  8                    & 12             & $\sim$3.5  \\ 
 \hline
\textbf{This work}          &  Parametric      &    \PthMin        &     \ILMin          &    \FinOperatingMHz&   \ISMaxLessThanPmax               & \ISMaxAtThirthydBm       & \PCBAreacm\\ 
 \hline
\end{tabular}
\label{Comparison_table}
\end{table*}

\subsection{Threshold and frequency reconfigurability}
After characterizing the operation of the built $pFSL$ at the $V_{DC}$ value (\Vbias V) resulting into the lowest $P_{th}$ around \FinOperatingApproxGHz GHz, we investigated the possibility to leverage different biasing conditions for the diode to reconfigure $P_{th}$ and the frequency at which the highest $IS$ value is desired. Yet, to ensure that the built $pFSL$ can be practically used to suppress EMI with different frequencies or power levels from the originally targeted value, we measured $P_{th}$, $IL^{s.s}$ and the maximum $IS$ for $P_{in}<P_{max}$ ($IS_{max}^{<P_{max}}$) for a broad range of $V_{DC}$ and $f_{in}$ values. This allowed us to construct three corresponding contour plots (see Fig.~\ref{Measured_Contour_Plots}) capturing the value of each performance metric for the analyzed $f_{in}$ and $V_{DC}$ values. As evident, through the strategic adoption of $V_{DC}$, the measured $pFSL$ can simultaneously achieve $P_{th}$ and $IL^{s.s}$ values lower than 2 dBm and 2 dB respectively, $IS_{max}^{<P_{max}}$ values up to 7 dB and a tunable operational frequency ranging from \FinOperatingRangeMinGHz GHz to \FinOperatingRangeMaxGHz GHz.

 \subsection{Comparison with the State-of-the-Art}
 
 To benchmark the performance attained by our $pFSL$ prototype with those attained by other passive diode-based FSLs, we compared (see Table \ref{Comparison_table}) the $P_{th}$, $IL^{s.s}$, $f_{in}^{opt}$ and $IS_{max}^{<P_{max}}$ of our built $pFSL$ when $V_{DC}$ is chosen to minimize $P_{th}$ with the corresponding values exhibited by the most recent demonstrated counterparts. Also, the $IS$ value ($IS^{28dBm}$) of the reported $pFSL$ for a much larger $P_{in}$ value (\PinSweepMax dBm) than $P_{th}$ was also compared to those of the other counterparts listed in Table \ref{Comparison_table} to assess the capability to protect any cascaded electronic components even from exceptionally strong EMI. As evident from Table \ref{Comparison_table}, the $pFSL$ reported in this work exhibits the lowest $P_{th}$ and $IL^{s.s}$ values among all the previously demonstrated diode-based passive FSLs, even though it is operating at one of the highest frequencies. In particular, the $P_{th}$ value attained by our reported $pFSL$ is nearly five times lower than what achieved by the previously reported SoA $pFSL$ counterparts operating within the same frequency range. Nevertheless, when $V_{DC}$ is selected to minimize $P_{th}$, $IS_{max}^{<P_{max}}$ is lower than what shown by the other previously reported diode-based FSLs. Yet, as shown in Fig.~\ref{Measured_Contour_Plots}, $IS_{max}^{<P_{max}}$ can be increased up to 7 dB by relying on slightly different $V_{DC}$ values, at the cost of  higher $P_{th}$ (still lower than 2 dBm) and $IL^{s.s}$ (still lower than 2 dB). Furthermore, it is important to point out that, differently from other listed prototypes, the reported $pFSL$ exhibits a large $IS^{28dBm}$ exceeding \ISMaxAtThirthydBm dB, limiting the maximum output power delivered to any cascaded components to \PMaxdBm dBm even in those scenarios when strong EMI with power approaching 30 dBm is received. Finally, by looking at the reported area of all the FSLs listed in Table \ref{Comparison_table}, it is easy to notice that our reported $pFSL$ shows the highest degree of miniaturization. 
 
 In the next section an alternative approach not relying on changes of $V_{DC}$ is introduced and experimentally validated to enhance both $IS_{max}^{<P_{max}}$ and $IS^{28dBm}$ without significantly degrading $P_{th}$ and while preserving record-low $IL^{s.s}$ values ($<$2 dB).

 \begin{figure}[p]
  \begin{center}
\begin{subfigure}{1.0\linewidth}
    \centering
    \caption{}
    \includegraphics[width=1.0\linewidth]{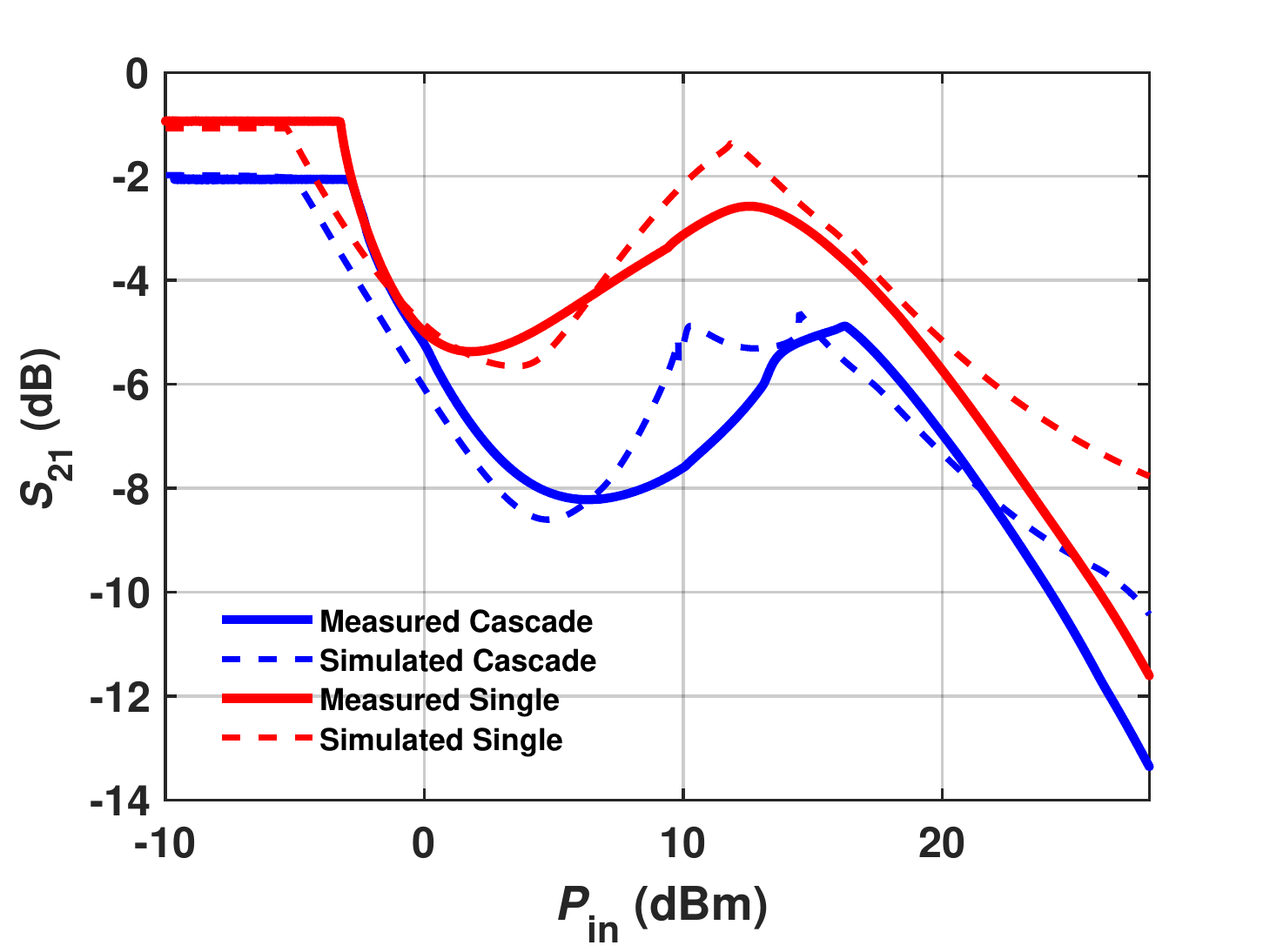} 
    \label{S21_Vs_Power_MeasVsSim_SingleVsCascade}
\end{subfigure}
\begin{subfigure}{1.0\linewidth}
    \centering
    \caption{}
    \def\big{\includegraphics[width=1.0\linewidth]{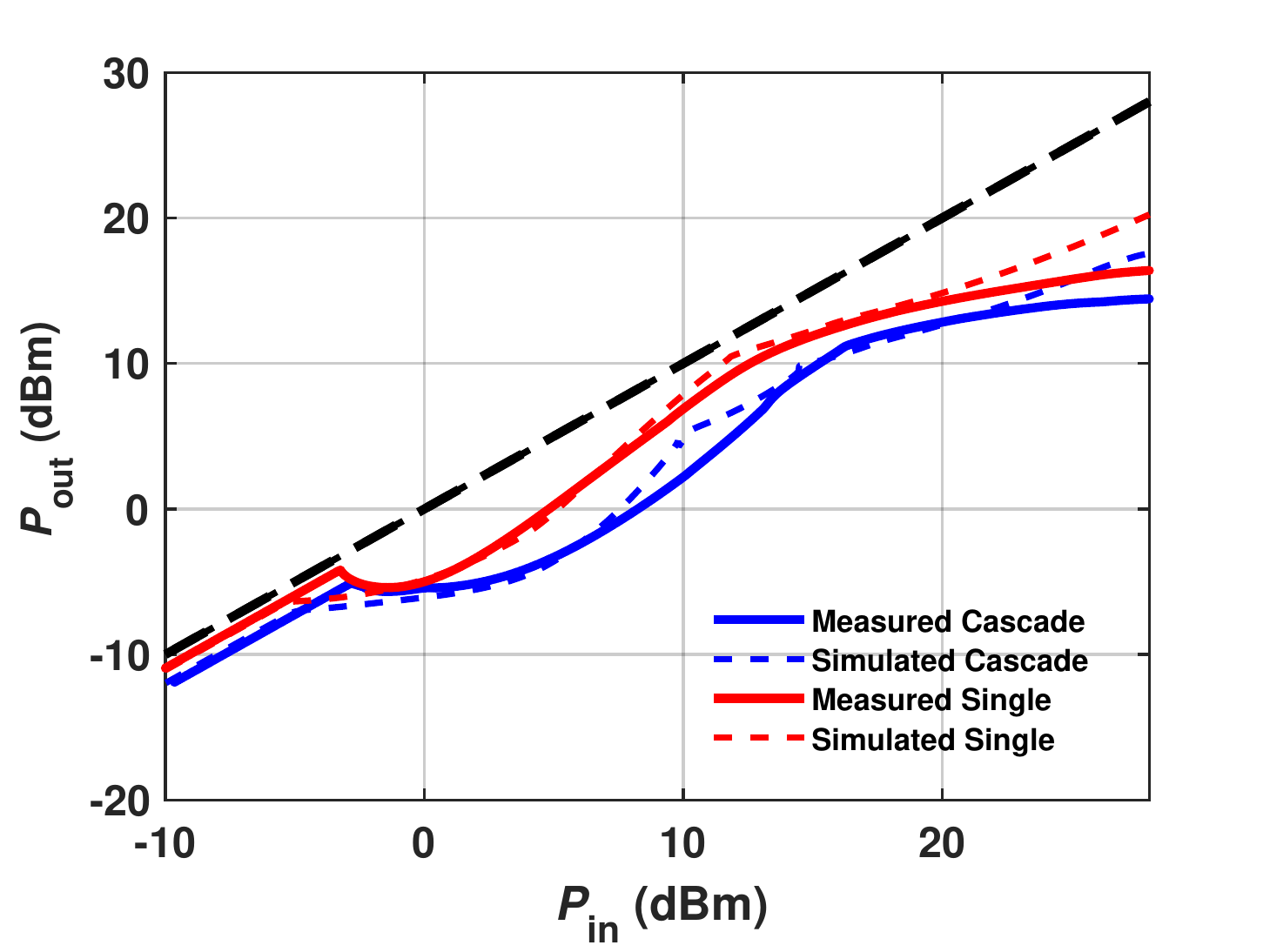}}  
    \def\little{\includegraphics[width=1.7in]{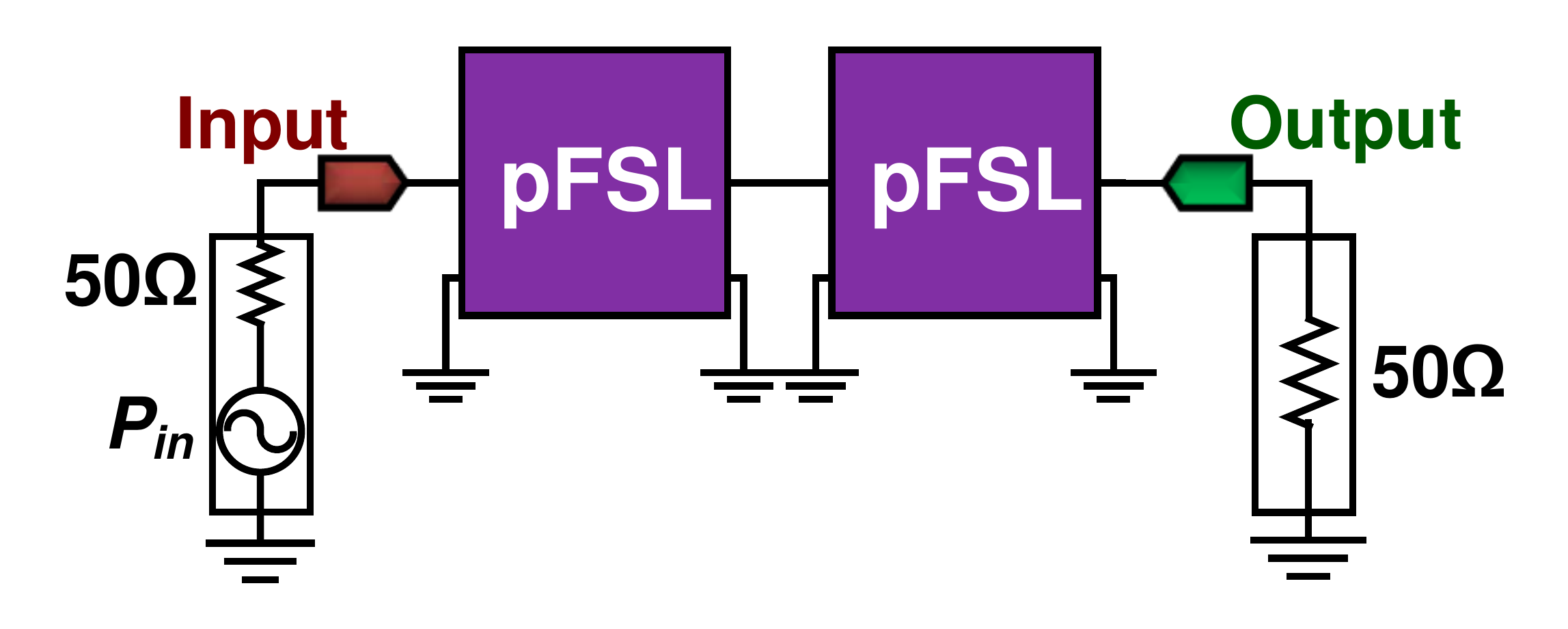}}
    \def\stackalignment{l}  
    \topinset{\little}{\big}{20pt}{34pt}   
    \label{Pout_Vs_Power_MeasVsSim_SingleVsCascade}
\end{subfigure}
\caption{ Measured (continuous lines) and simulated (dotted lines) trends of the $S_{21}$ (a) and of the output power (b) for $P_{in}$ values ranging from -\PinSweepMin dBm to \PinSweepMax dBm and when using only one $pFSL$ ($i.e.$ M=1, in red) or two identical $pFSLs$ ($i.e.$ M=2, in blue). A schematic representation of the connection between the $pFSLs$ for M=2 is shown in the inset. 
}\label{MeasVsSim_SingleVsCascade}
  \end{center}
\end{figure}


 \subsection{Increasing $IS$ through multiple $pFSL$ stages}
Similarly to what was previously shown for absorptive $pFSLs$\cite{Ramirez2008}, cascading multiple reflective $pFSLs$ provides useful means to increase $IS_{max}^{<P_{max}}$, augmenting the maximum achievable suppression at $f_{in}^{opt}$. Nevertheless, this technique can be practically leveraged only when absorptive or reflective $pFSLs$ with low $IL^{s.s}$ are available, such as the one we designed and built in this work. In fact, since the insertion-loss of a chain of $pFSLs$ ($IL_{chain}^{s.s}$) grows proportionally with the number of cascaded stages ($M$), there exists an inevitable trade-off between the maximum exploitable $M$ and the highest tolerated $IL_{chain}^{s.s}$. Also, differently from any chains of absorptive $pFSLs$ whose design and operation inevitably lead to $P_{th}$ values increasing proportionally to $M$, the high $Z_{in}$ (see Fig.~\ref{generic_schematic}) value exhibited by reflective $pFSLs$ for $P_{in}<P_{th}$ renders the voltage at $f_{in}^{opt}$ across all the adopted diodes almost independent of $M$, especially when $Z_{tx}$ is chosen to be much higher than $R_s$ in order to minimize $IL^{s.s}$. This key operational feature allows to preserve low $P_{th}$ values even when multiple reflective $pFSL$ stages are used to enable higher $IS_{max}^{<P_{max}}$. Moreover, contrary to absorptive $pFSLs$, the adoption of multiple reflective $pFSL$ stages permits to increase even the $IS$ values attained for much higher $P_{in}$ values than $P_{max}$. 

In order to demonstrate the capability to achieve higher $IS_{max}^{<P_{max}}$ values through the adoption of multiple $pFSL$ stages, we built a copy of the $pFSL$ discussed in the previous section. The two $pFSLs$ were then connected to each other and the modified trends of the $S_{21}$ $vs.$ $P_{in}$ were extracted (Fig.~\ref{S21_Vs_Power_MeasVsSim_SingleVsCascade}), along with the corresponding trend of the output power $vs.$ $P_{in}$ at \FinOperatingGHz GHz ($i.e.$, the $f_{in}$ value giving the lowest threshold for a single stage $pFSL$). As evident from Fig.~\ref{MeasVsSim_SingleVsCascade}, the chain formed by the two built $pFSL$ stages allows to significantly enhance the maximum $IS_{max}^{<P_{max}}$ value attained by just one stage, while causing negligible ($<$1 dB) increases of  $P_{th}$ and $IL^{s.s}$.


%
%
%
\section{Conclusion}
In this article, we discussed the design criteria and measured performance of a $\sim$\FinOperatingApproxGHz GHz diode-based reflective parametric frequency selective limiter ($pFSL$) built on a FR-4 printed-circuit-board (PCB) and using commercial off-the-shelf components. Thanks to its engineered dynamics, the reported $pFSL$ prototype can exhibit record-low insertion-loss for low-power signals ($IL^{s.s}$, as low as \ILMin dB), record-low power threshold ($P_{th}$, as low as \PthMin dBm) and a significant suppression (up to \ISMaxLessThanPmax dB) for input power levels lower than the one forcing the diode to operate in its forward conduction. Furthermore, due to its unique design characteristics and regardless of the inevitable reduction in frequency selectivity, the built $pFSL$ ensures a good protection even from much stronger interference signals with power approaching \PinSweepMax dBm. In addition, by strategically tuning the DC-biasing voltage of the diode, the reported $pFSL$ allows to reconfigure the frequency at which the maximum $IS$ is obtained by nearly \FreqOperatingRangeMHz MHz (corresponding to a tuning range of $\sim$\FreqTuningRange), while simultaneously preserving low $P_{th}$ ($<$2 dBm) and $IL^{s.s}$ ($<$ 2 dB) values. Finally, by connecting two copies of the same $pFSL$ designed and built in this work, we demonstrated that a significantly larger suppression value ($>$8 dB) for high-power signals can be attained, while preserving low $P_{th}$ ($<$-2.5 dBm) and low $IL^{s.s}$ ($<$2 dB).

%
%
\section*{Acknowledgment}
This work has been funded by the National Science Foundation (NSF) under award \#1854573.
%
\bibliographystyle{IEEEtran}
\bibliography{IEEEabrv,Bibliography}




\begin{IEEEbiography}[{\includegraphics[width=1in,height=1.25in,clip,keepaspectratio]{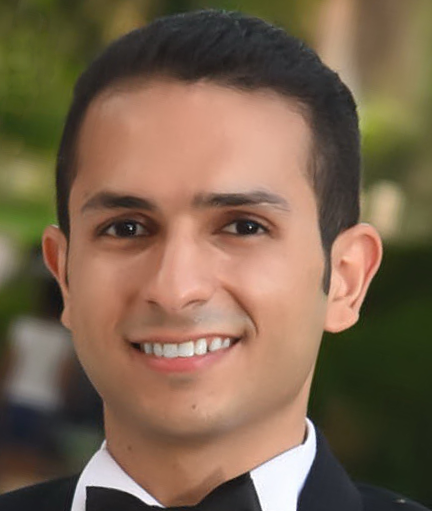}}]{Hussein M. E. Hussein}
(S’20) received his B.S. and M.Sc. in electrical engineering at Cairo University, Giza, Egypt, in  2013  and  2017,  respectively. He is  currently  pursuing the  Ph.D. degree with the  Electrical  and  Computer  Engineering  department at Northeastern University, Boston, MA, USA. He is currently working on the development of parametric phase noise reduction techniques, for RF systems, based on nonlinear devices and circuits.
\end{IEEEbiography}
\vskip -20pt plus -1fil
\begin{IEEEbiography}[{\includegraphics[width=1in,height=1.25in,clip,keepaspectratio]{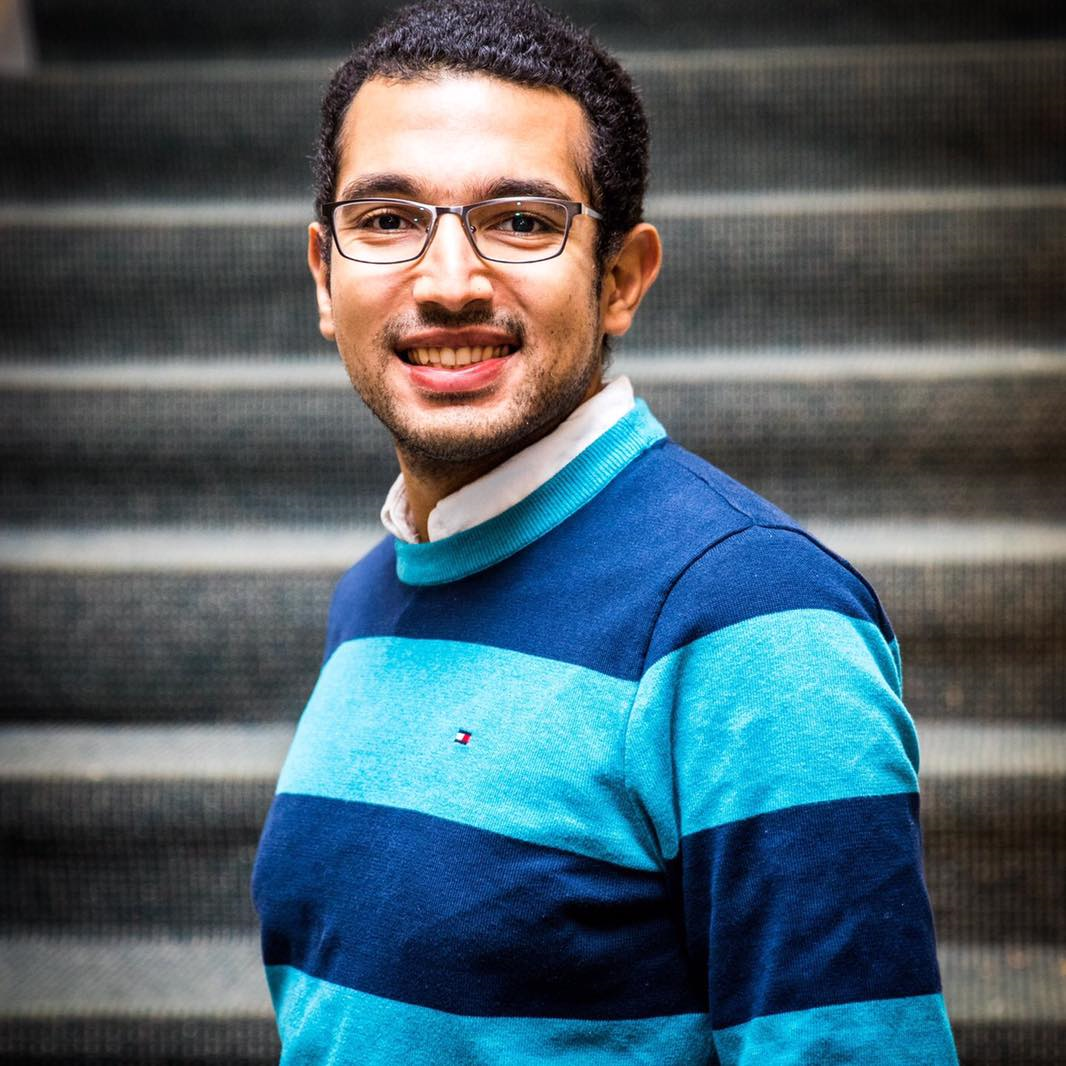}}]{Mahmoud A. A. Ibrahim}
(S’13) received the B.Sc. (Hons.) and M.Sc. degrees in electrical engineering from the Electronics and Electrical Communications Engineering Department, Cairo University, Giza, Egypt, in 2013 and 2015, respectively. He is currently pursuing the Ph.D. degree in electrical engineering with Northeastern University, Boston, MA, USA.\\
From 2013 to 2015, he was a Teaching and Research Assistant with Cairo University. In the summer of 2018, he joined the PLL Team at Qualcomm, San Diego, CA, USA, as an Analog-Mixed Signal Design Intern; where he was involved in the research and design of ultra-low power multi-Giga Hertz LC-oscillators in deep-submicron Fin-FET technologies. Since 2016, he has been a Graduate Research and a Teaching Assistant with Northeastern University. He is also working on the design of ultra-low power transceivers for biomedical applications. His research interests include integrated analog, mixed-signal, RF circuits for low-power wireless transceivers, and power management integrated circuits.
\end{IEEEbiography}
\vskip -35pt plus -1fil

%
\begin{IEEEbiography}[{\includegraphics[width=1in,height=1.25in,clip,keepaspectratio]{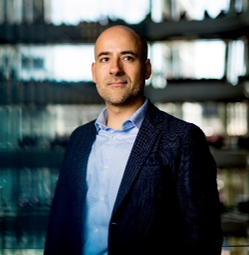}}]{Matteo Rinaldi}
is a Professor in the Electrical and Computer Engineering department at Northeastern University and the Director of Northeastern SMART a university research center that, by fostering partnership between university, industry and government stakeholders, aims to conceive and pilot disruptive technological innovation in devices and systems capable of addressing fundamental technology gaps in several fields including the Internet of Things (IoT), 5G, Quantum Engineering, Digital Agriculture, Robotics and Healthcare. Dr. Rinaldi received his Ph.D. degree in Electrical and Systems Engineering from the University of Pennsylvania in December 2010. He worked as a Postdoctoral Researcher at the University of Pennsylvania in 2011 and he joined the Electrical and Computer Engineering department at Northeastern University as an Assistant Professor in January 2012. Dr. Rinaldi’s group has been actively working on experimental research topics and practical applications to ultra-low power MEMS/NEMS sensors (infrared, magnetic, chemical and biological), plasmonic micro and nano electromechanical devices, medical micro systems and implantable micro devices for intra-body networks, reconfigurable radio frequency devices and systems, phase change material switches, 2D material enabled micro and nano mechanical devices. 
The research in Dr. Rinaldi’s group is supported by several Federal grants (including DARPA, ARPA-E, NSF, DHS), the Bill and Melinda Gates Foundation and the Keck Foundation with funding of \$14+M since 2012. 
Dr. Rinaldi has co-authored more than 140 publications in the aforementioned research areas and also holds 10 patents and more than 10 device patent applications in the field of MEMS/NEMS. 
Dr. Rinaldi was the recipient of the IEEE Sensors Council Early Career Award in 2015, the NSF CAREER Award in 2014 and the DARPA Young Faculty Award class of 2012. He received the Best Student Paper Award at the 2009, 2011, 2015 (with his student) and 2017 (with his student) IEEE International Frequency Control Symposiums; the Outstanding Paper Award at the 18th International Conference on Solid-State Sensors, Actuators and Microsystems, Transducers 2015 (with his student) and the Outstanding Paper Award at the 32nd IEEE International Conference on Micro Electro Mechanical Systems, MEMS 2019 (with his student).
Prof. Rinaldi is the founder and CEO of Zepsor Technologies, a start-up company that aims to bring to market zero standby power sensors for various internet of things applications including distributed wireless fire monitoring systems, battery-less infrared sensor tags for occupancy sensing and distributed wireless monitoring systems of plant health parameters for digital agriculture.
Prof. Rinaldi is also the owner of Smart MicroTech Consulting LLC, a company that routinely provides consulting services to government agencies, large companies and startups in the broad areas of Micro and Nano Technologies, Internet of Things, Wireless Communication devices and systems, Radio Frequency Devices and Systems and Sensors.  
\end{IEEEbiography}

\begin{IEEEbiography}[{\includegraphics[width=1in,height=1.25in,clip,keepaspectratio]{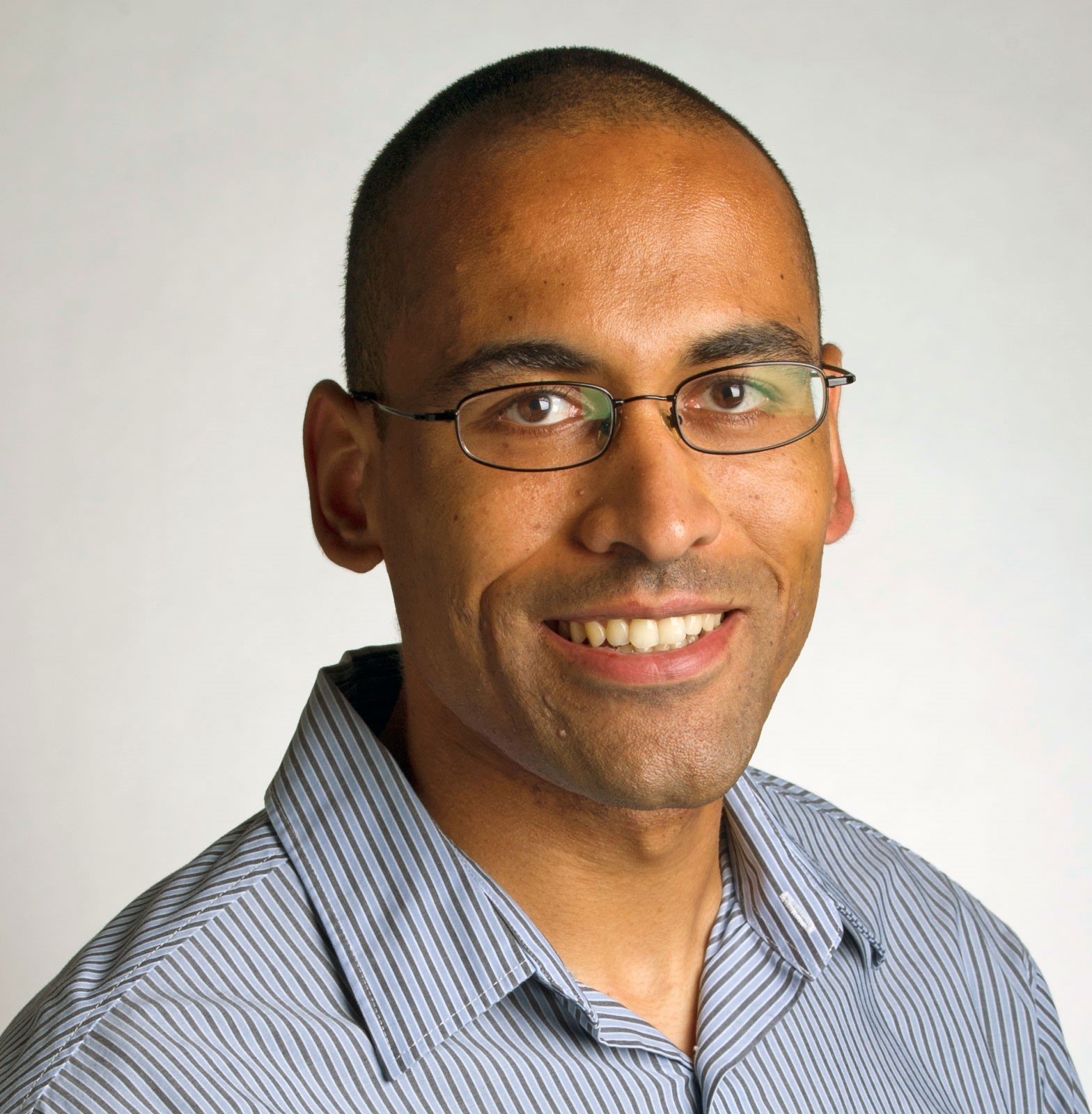}}]{Marvin Onabajo}
(S’01–M’10–SM’14) is an Associate Professor in the Electrical and Computer Engineering Department at Northeastern University. He received a B.S. degree (summa cum laude) in Electrical Engineering from The University of Texas at Arlington in 2003, as well as the M.S. and Ph.D. degrees in Electrical Engineering from Texas A\&M University in 2007 and 2011, respectively. 
From 2004 to 2005, he was Electrical Test/Product Engineer at Intel Corp. in Hillsboro, Oregon. He joined the Analog and Mixed-Signal Center at Texas A\&M University in 2005, where he was engaged in research projects involving analog built-in testing, data converters, and on-chip temperature sensors for thermal monitoring. In the spring 2011 semester, he worked as a Design Engineering Intern in the Broadband RF/Tuner Development group at Broadcom Corp. in Irvine, California. Marvin Onabajo has been at Northeastern University since the Fall 2011 semester. His research areas are analog/RF integrated circuit design, on-chip built-in testing and calibration, mixed-signal integrated circuits for medical applications, data converters, and on-chip sensors for thermal monitoring. He currently serves as Associate Editor on the editorial boards of the IEEE Transactions on Circuits and Systems I (TCAS-I, 2016-2017, 2018-2019, and 2020-2021 terms) and of the IEEE Circuits and Systems Magazine (2016-2017, 2018-2019, and 2020-2021 terms). During the 2014-2015 term, he was on the editorial board of the IEEE Transactions on Circuits and Systems II (TCAS-II). He received a 2015 CAREER Award from the National Science Foundation, a 2017 Young Investigator Program Award from the Army Research Office (ARO), and the 2015 Martin Essigman Outstanding Teaching Award from the College of Engineering at Northeastern University.
\end{IEEEbiography}
%
\begin{IEEEbiography}[{\includegraphics[width=1in,height=1.25in,clip,keepaspectratio]{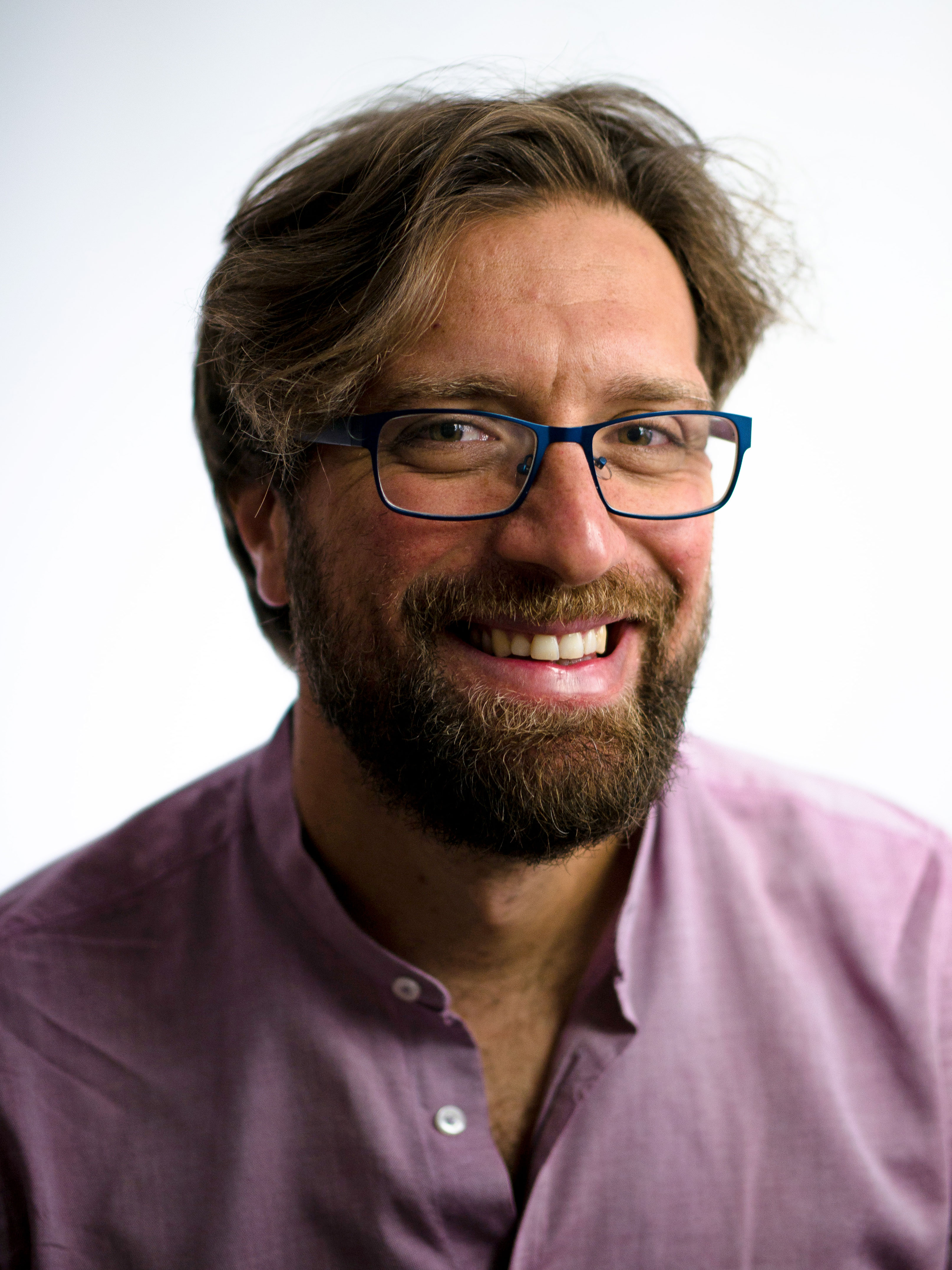}}]{Cristian Cassella} is an Assistant Professor in the Electrical and Computer Engineering department at Northeastern University, Boston (USA). He received his B.S.E. and M.Sc., with honors, at University of Rome – Torvergata in 2006 and 2009, respectively. In 2011, he was a visiting scholar at University of Pennsylvania. In 2012 he entered a Ph.D. program at Carnegie Mellon University which he completed in 2015. In 2015 he was a Postdoctoral Research Associate at Northeastern University. In 2016, he became Associate Research Scientist. He is author of 80 publications in peer-reviewed journals and conference proceedings. Two of his peer-reviewed journal papers published on the IEEE Journal of MicroElectroMechanical systems (JMEMS) were selected as papers of excellent quality (JMEMS RightNowPapers), hence being released as open-access. One of his journal papers was chosen as the cover for the Nature Nanotechnology October 2017 issue. An other one of his journal papers was selected as a featured article by the Applied Physics Letters magazine. He won the best paper award at the IEEE International Frequency Control Symposium (2013, Prague). In 2018, he was awarded by the European Community (EU) the Marie-Sklodowska-Curie Individual Fellowship. He holds four patents and four patent applications in the area of acoustic resonators and RF systems. He is a technical reviewer for several journals, such as Applied Physics Letter, IEEE Transactions on Electron Devices, IEEE Transactions on Ultrasound, Ferroelectric and Frequency Control, IEEE Journal of MicroElectroMechanical devices, IEEE Electron Device Letter, Journal of Micromachine and Micro-Engineering, Journal of Applied Physics, IEEE Sensors Letter and Review of Scientific Instruments.
\end{IEEEbiography}

\end{document}